\documentclass[pre,twocolumn,showpacs,preprintnumbers,amsmath,amssymb,superscriptaddress]{revtex4-2}

\usepackage{graphicx}
\usepackage{dcolumn}
\usepackage{bm}
\usepackage{color}
\usepackage{hyperref}
\usepackage{amsmath}
\usepackage[capitalise]{cleveref}
\usepackage{amsfonts}
\usepackage{amsthm}
\usepackage{cases}
\usepackage{amssymb}
\usepackage{latexsym}
\usepackage{multirow}
\usepackage{xcolor}
\usepackage{mathrsfs}
\usepackage{textcomp}
\usepackage{lipsum}
\hypersetup{colorlinks=true,linkcolor=blue,citecolor=blue}

\newcommand{\ve}[1][K]{\mathbf{#1}}

\newcommand{\dd}{\mbox{d}}

\begin{document}

\title{Effective description of Taylor dispersion in strongly corrugated channels}

\author{Arthur Alexandre}
\affiliation{Univ. Bordeaux, CNRS, LOMA, UMR 5798, F-33400, Talence, France.}
\affiliation{ Institute of Bioengineering, School of Life Sciences, \'{E}cole Polytechnique F\'{e}d\'{e}rale de Lausanne (EPFL), CH-1015 Lausanne, Switzerland.}
\affiliation{SIB Swiss Institute of Bioinformatics, CH-1015 Lausanne, Switzerland.}
 \author{Thomas Gu\'{e}rin}\email{thomas.guerin@u-bordeaux.fr}
\affiliation{Univ. Bordeaux, CNRS, LOMA, UMR 5798, F-33400, Talence, France.}
\author{David S. Dean} \email{david.dean@u-bordeaux.fr}
\affiliation{Univ. Bordeaux, CNRS, LOMA, UMR 5798, F-33400, Talence, France.}

\begin{abstract}
We consider Taylor dispersion in periodic but highly corrugated channels. Exact analytical expressions for the long-time diffusion constant and drift along the channel are derived to next-to-leading order in the limit of small channel period. Using these results, we show how an effective model for Taylor dispersion in \textcolor{black}{ porous media with tortuous pores}
can be framed in terms of dispersion in a uniform channel with absorption/desorption at its surface, an effective slip length for the flow at the surface and an effective, universal, diffusion constant on the surface. This work thus extends the concept of an effective slip-length for hydrodynamics flows to Taylor dispersion  by those flows. The analytical results are confirmed by numerical calculations, and present a robust method to understand and upscale  the transport properties of flows in \textcolor{black}{pore like geometries}.
\end{abstract}

\maketitle

\section{Introduction}

The transport properties of tracer particles, such as pollutants or reactants, in heterogeneous media at large temporal and spatial scales play a vital role  in  fluid mechanics, hydrology, chemical engineering, soft matter and solid state physics \cite{marbach2018transport,aminian2016boundaries,kim2019tuning,brenner1993macrotransport,meigel2022dispersive}. 
The average velocity (or drift) and late time diffusivity are essential quantities in the  study of mixing \cite{le2013stretching,dentz2011mixing,de2012flow}, sorting \cite{bernate2012stochastic}, chemical delivery \cite{aminian2016boundaries,aminian2015squaring}   and chemical reaction dynamics \cite{dentz2011mixing,brenner1993macrotransport}. 
In heterogeneous media, spatial variations in the local diffusivity and drift fields can lead to drastic differences between effective transport coefficients and microscopic ones \cite{dean2007effective}. For instance, in confined media,  composed of hard obstacles with reflecting walls and in the absence of background hydrodynamic flows, the diffusivity is  smaller (sometimes drastically smaller) than the microscopic diffusivity, due to entropic trapping effects~\cite{malgaretti2013entropic,burada2009diffusion,jacobs1967diffusion,reguera2006entropic,rubi2019entropic}. However, in the presence of spatially varying hydrodynamic flows, the long-time diffusivity is much larger than the microscopic one, due to the difference in histories of sampling the velocity field: the  phenomenon of Taylor dispersion~\cite{taylor1953dispersion}. 

\begin{figure}
\includegraphics[width=\linewidth]{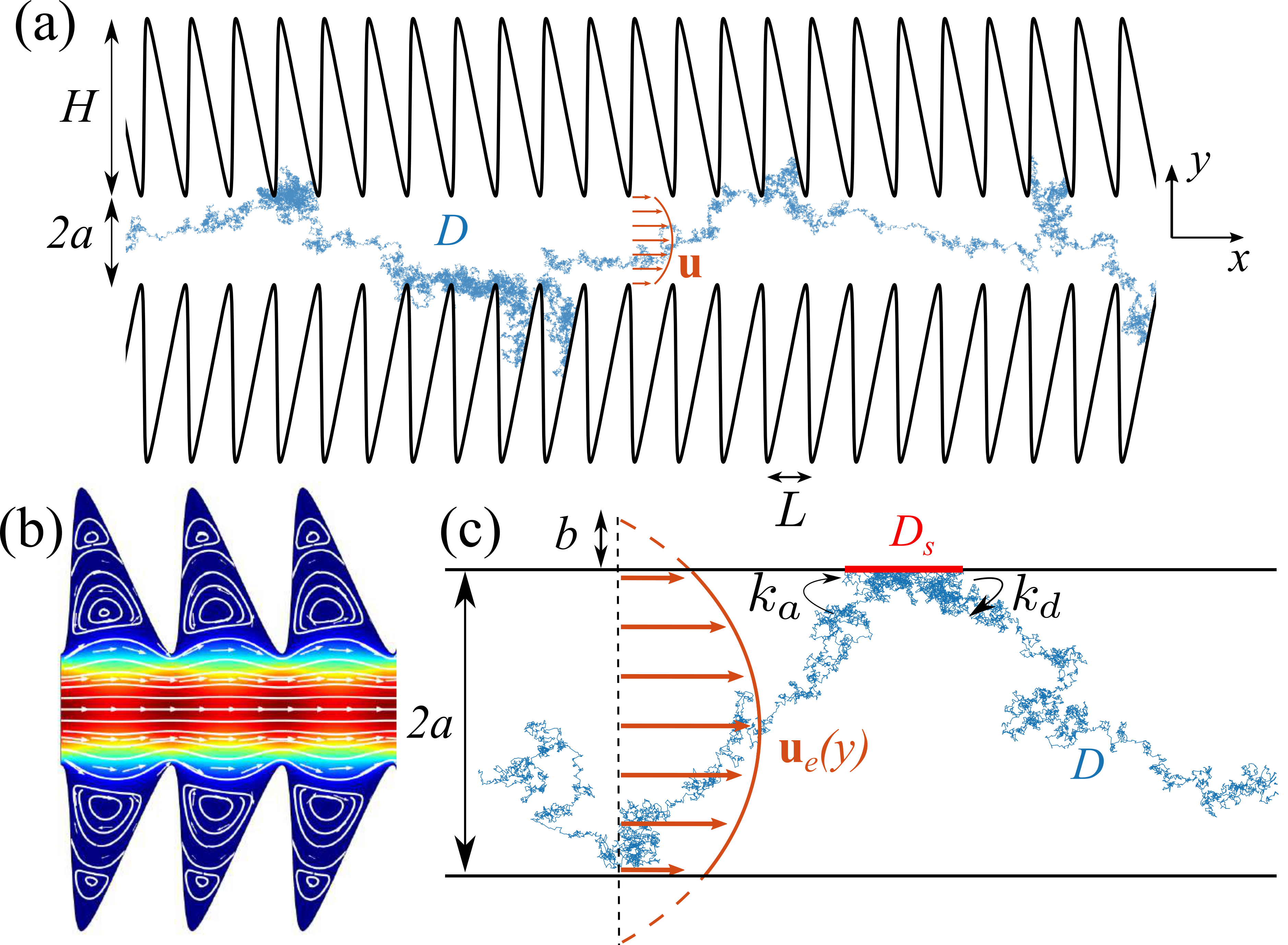}
\caption{(a) Schematic of  particle diffusion in an highly corrugated channel advected by a pressure induced hydrodynamic flow $\ve[u]$.  (b) Hydrodynamic flow in this channel, for $L/a = 2$ and $H/a = 3$.   
Colors represent the norm of the velocity field $|\ve[u]|$ (from low value in blue to high values in red, in arbitrary units), white lines are the stream lines. (c) Simplified problem of a particle diffusing in a channel of uniform width, with reversible binding at the walls, surface diffusivity $D_s$ and a flow calculated with a slip length $b$. Here we show how to map the situation (a) onto the transport problem in uniform channels exactly at first order when $L/a\to0$. }
\label{fig:Fig1}
\end{figure} 

In the presence of both obstacles and flows, such as in the case of steady state (so non-turbulent) pressure induced flows in non-uniform channels, the effects of  entropic trapping  and  dispersion (by non-uniform flows) compete in a non-trivial way. Most analytical results for this problem have been derived for slowly varying channels \cite{martens2013hydrodynamically,kalinay2020taylor,bolster2009solute,yang2017hydrodynamic,smith1983longitudinal,hoagland1985taylor,marbach2019active,dagdug2012projection} or with perturbative approaches for nearly flat surfaces \cite{roggeveen2023transport,marbach2018transport}. The fact that the flow itself depends on the geometry of obstacles means that for  complex geometries it is often necessary to treat the problem numerically \cite{haugerud2022solute,amaral1997dispersion_part1,amaral1997dispersion_part2,quintard1994convection,quintard1993transport}. In hydrodynamics, for a flow in the vicinity of a structured surface presenting protrusions  \cite{richardson1973no,jeong2001slip,jansons1988determination} or variations in surface properties, a very useful concept is that of an effective \text{slip length} \cite{lauga2005microfluidics,bocquet2010nanofluidics,squires2005microfluidics} which takes into account, in a coarse-grained way, the microscopic aspects of the surface. 
A similar, but much less studied, question arises as to whether one can define effective boundary conditions for the advection-diffusion of particles moving near a surface presenting protrusions as shown in Figs.~\ref{fig:Fig1}(a),(b).  
 Clearly, when trapped in a protrusion  tracer dispersion parallel to the channel is reduced. It is natural to ask if this reduction can be captured using a model of a uniform channel with a sticky surface, that is to say a  surface having absorption/desorption of particles (corresponding to entering/leaving a protrusion), as well as an effective surface diffusion constant (modeling the fact that there is still some dispersion along the channel due to the finite width of the protrusions).
This effective picture was introduced in the context of diffusion in comb-like geometries \cite{benAvraham2000}, for example in Refs.~\cite{dagdug2014aris,berezhkovskii2014normal} to study porcuepine-like  cylindrical channels connected to very thin  finite tubes of finite volume (thus taking up little of the surface area of the tube) and  in \cite{bettarini2024effective} for rectangular protrusions, without flow. 
However, for channels with a generic shape, for which protrusions are not necessarily thin, the mapping onto a flat sticky channel is not obvious since particles in the entrance of the protrusion zones are clearly not completely immobile nor completely mobile. 

Here, we show how, in the limit of small period, an effective model can be derived which takes into account this effect, and we derive the precise parameters of the effective model depend on the geometry of the protrusions. This means that, in a way similar to the description of  the effective electrical properties  of complex circuits in terms of simpler ones, the transport in this  problem can be described by a cylindrical (or planar) geometry [Fig.~\ref{fig:Fig1}(c)] with effective surface absorption/desorption properties, an effective slip length for the flow, and an effective surface diffusion constant. 
As well as its underlying physical interest, the theory here clearly has potential to be used for upscaling methods for transport in tortuous porous media, \textcolor{black}{where the effect of dead ends (in which the flow cannot penetrate) could be taken into account via effective (and potentially spatially varying) adsorption and desorption coefficients as well as effective surface diffusivity. }

\textcolor{black}{The outline of the paper is as follows. In Section \ref{SecPhysPb}, we introduce the physical problem of diffusion in channels and our formalism to calculate the diffusivity. In Section \ref{SectionMatchedAsympt}, we provide the solution of this problem in the strong corrugation limit. In Section \ref{SecUniform}, we compare our results to the case of dispersion in uniform channels, with flat but sticky walls, and we show how to identify analytically the effective adsorption and desorption coefficients, the effective (hydrodynamic) slip length and the effective surface diffusivity. Our results are validated numerically in Section \ref{SecNum} and generalized to 3D axisymmetric channels in Section \ref{Gen3D}. }

\section{Physical problem} 
\label{SecPhysPb}
Consider a point-like tracer particle diffusing with microscopic isotropic diffusivity $D$  and advected by incompressible flow $\ve[u](\ve[r])$, in a pore like geometry with rapid undulations perpendicular to the channel axis. Specifically, we assume that geometry is a channel of width $2h(x)$  in $d=2$ for symmetric channel, or of local radius $h(x)$ in $d=3$ for axisymmetric tubes, where $x$ here denotes the coordinate along the channel. We assume that $h$ and $\ve[u]$ are periodic in the $x$ direction, with period $L$. This periodicity ensures that at long times one can define effective drift $v$  and diffusivity $D_e$ as 
\begin{align}
v=\frac{\overline{[x(t)-x(0)]}}{t},\
 D_e\underset{t\to\infty}{\simeq}\frac{\overline{[x(t)-x(0)-v t]^2}}{2t},
\end{align}
where $\overline{\cdots}$ denotes the ensemble average over stochastic trajectories. The average local drift is  $v=\Omega^{-1}\int_\Omega \dd\ve[r]\  u_x(\ve[r])$, with $\Omega$ the volume of one period of the channel. To calculate the effective diffusivity, many approaches have been proposed, such as the homogenization approach \cite{rubinstein1986dispersion,quintard1994convection,alshare2010modeling,amaral1997dispersion_part1,amaral1997dispersion_part2}, macrotransport theory \cite{brenner1980dispersion,brenner1993macrotransport}, or Kubo-type formulas \cite{guerin2015kubo,guerin2015force}, where $D_e$ is expressed as 
\begin{align}
D_e = D-\int_0^\infty \dd t \ \overline{[V_x(\ve[r](t))-\overline{V_x}][V_x^*(\ve[r](0))-\overline{ V_x^*}]}\label{eq:Kubof}.
\end{align}
Here, $\ve[V]=\ve[u]+D\,\ve[n]\,\delta_s(\ve[r])$ is the local drift, where  $\delta_s$ is the surface delta function  formally representing the ``kicks'' applied to the particle when it touches the reflecting surface, preventing it from entering the obstacles \textcolor{black}{(see Appendix \ref{sec:SecFormalism})}. Concretely, the term $\ve[n]$ is the normal to the surface (oriented towards the fluid)  and $\delta_s$ defined such that   $\int_\Omega \dd \ve[r]\,\delta_s(\ve[r]) g(\ve[r])=\int_{\partial \Omega} \dd S\, g(\ve[r])$ for any function $g$. Next, $\ve[V]^*$ is the drift \textit{after time reversal}, which is equal to  $\ve[V]^*=-\ve[u](\ve[r])+D\,\delta_s(\ve[r])\,\ve[n]$ since in the time-reversed situation, particles see the reversed flow field $-\ve[u]$ and are submitted to the same obstacles. The correlation function in \cref{eq:Kubof} can then  be calculated by defining an auxiliary field $f$ as
\begin{align}
f(\ve[r])=-\int_0^\infty \dd t \int_\Omega \dd \ve[r]_0  \ P(\ve[r],t \vert \ve[r]_0 ) [ V_x^*(\ve[r]_0)- \overline{V_x^*}]\label{eq:Def_f},
\end{align}
where $P(\ve[r],t \vert \ve[r]_0)$ is the probability density of observing $\ve[r]$ (modulo the period) at $t$, starting from $\ve[r]_0$. This probability density \textcolor{black}{satisfies the Fokker-Planck equation
\begin{align}
&\partial_t P = -\nabla \cdot[\ve[u]\, P -D\nabla P], & \ve[n]\cdot[\ve[u]\, P -D\nabla P]_{\ve[r]\in\partial\Omega}=0\label{FkPEq}
\end{align}}and reaches a steady state which is uniform due to the incompressibility of the flow, $P(\ve[r],t\to\infty)=1/\Omega$.  With the definition \textcolor{black}{(\ref{eq:Def_f}) of $f$}, the diffusivity (\ref{eq:Kubof})  is given by
\begin{align}
 D_e& = D+ \frac{1}{\Omega}\int_\Omega \dd\ve[r] \ (u_x f-D\partial_x f )   \label{eq:ExprDe}.
\end{align}
\textcolor{black}{Furthermore, one can show that the auxiliary field $f(\ve[r])$  satisfies a partial differential equation. Clearly, using the Fokker-Planck equation (\ref{FkPEq}), one obtains
\begin{align}
-\nabla\cdot& [\ve[u]\, f(\ve[r])]+D\,\nabla^2 f(\ve[r])=\nonumber\\
&-\int_0^\infty \dd t \int_\Omega \dd \ve[r]_0\,   \partial_tP(\ve[r],t \vert \ve[r]_0 )\,  [ V_x^*(\ve[r]_0)- \overline{V_x^*}].
\end{align}
The integration over $t$ can be performed, leading to
\begin{align}
-\nabla\cdot&(\ve[u]f)+D\nabla^2f (\ve[r])\nonumber\\
&=-\int_\Omega \dd\ve[r]_0 \left[ \frac{1}{\Omega}-\delta(\ve[r]-\ve[r]_0)\right]  [ V_x^*(\ve[r]_0)- \overline{V_x^*}]\nonumber\\
&=-u_x(\ve[r])+\overline{u_x}+D\,\ve[n]\,\cdot \ve[e]_x \,\delta_s(\ve[r])\label{eq:942}.
\end{align}
If $\ve[r]$ is in the bulk, the delta-surface term vanishes and one obtains 
\begin{align}
&-\ve[u]\cdot\nabla f+D\nabla^2 f= v-u_x &(\ve[r]\in \Omega). \label{eq:EqForf}
\end{align}
To find the boundary condition for $f$, we argue that $P(\ve[r],t)$, and therefore $f$ vanishes for all $\ve[r]$ inside the obstacles. Therefore, if one integrates \cref{eq:942} over a small volume that includes a portion of the surface, taking into account the fact that  $\ve[u]$ vanishes at the boundary (the no-slip boundary condition), 
one finds directly 
\begin{align}
&D\ \ve[n]\cdot \nabla f =D \ \ve[n]\cdot\ve[e]_x  & (\ve[r]\in \partial \Omega),  \label{eq:BCForf}
\end{align}
 where $\ve[e]_x$ is the unit vector in the $x$ direction. }
 In addition, it is clear from \cref{eq:Def_f} that $f$ is periodic and must be of zero spatial average, $\int_\Omega \dd\ve[r] f(\ve[r])=0$.  

The flow field is assumed to be  a steady Stokes flow 
\begin{align}
&\nabla\cdot\ve[u]=0, & \eta\ \nabla^2 \ve[u]-\nabla \Pi =0, \label{eq:Flow}
\end{align}
where $\Pi$ the pressure field and $\eta$ the fluid viscosity. At the boundaries, the flow vanishes. Since the channel is periodic, the flow is also periodic, and so is the pressure gradient. As a consequence, we can define a parameter $(\nabla \Pi)_\infty$, which sets the magnitude of the flow, so that $\Pi(\ve[r]+L\ve[e]_x)=\Pi(\ve[r])+(\nabla \Pi)_\infty L$. 
While the case of weakly varying channels has been studied analytically at length \cite{martens2013hydrodynamically,kalinay2020taylor,bolster2009solute,yang2017hydrodynamic,smith1983longitudinal,hoagland1985taylor,marbach2019active,roggeveen2023transport,marbach2018transport}, here we will study the opposite limit $L\to0$ where all standard Fick-Jacobs and lubrication methods completely breakdown.
 
\section{Analytical solution for the strong corrugation limit} 
\label{SectionMatchedAsympt}
For fast varying channel $L/a\ll1$, where $a=\text{min}[h(x)]$, our method of solution relies on identifying the asymptotic behavior of $\ve[u]$ and $f$ in three different regions, and using the matched asymptotic expansion method. First, we define the central region (``$c$'') as the region where $\vert y\vert< a$, and the peripheral regions (``$p$'') as those with $a<\vert y\vert<h(x)$, corresponding to the locations in the lateral protrusions. Here $y$ is the distance to the central axis. We also define an \textit{inner} region that connects these peripheral and the central regions at the length scale $L$, which is small in the limit considered here. For the flow, we expect that in the central region it tends to a finite Poiseuille flow for vanishing $L$, while it vanishes in the lateral pores (as will be justified later), and it may be weak (of order $L/a$) in the inner region, so that its structure reads
\begin{align}
& \ve[u] \underset{L\to0}{\simeq}
\begin{cases}
\ve[u]_0(X,y)+L \ve[u]_1(X,y)+...&[y<a],\\
L\ \ve[u]^*(X,Y)+...&[Y=\mathcal{O}(1)],\\
0 &[y>a], 
\end{cases}\label{eq:ExpFlow}
\end{align}
where we have defined the rescaled coordinates 
\begin{align}
&X=x/L, & Y=(y-a)/L. 
\end{align}
For the function $f$, we  assume  the following series expansion in each region
\begin{align}
&f=\begin{cases}
\sum_{n\ge0}L^n f_n^c(X,y)& [y<a],\\
f_0^*(X,Y)+Lf_1^*(X,Y)+... & [Y=\mathcal{O}(1)],\\
\sum_{n\ge0}L^n f_n^p(X,y) & [y>a].
\end{cases}
\end{align}
The solutions in each region are found by inserting the above \textit{ansätze} into the equations for $f$ and $\ve[u]$, solving the resulting equations, and requiring that the solutions can be matched. For example, we require that $f_0^*(X,Y\to\infty)=f_0^p(X,a)$, $f_0^*(X,Y\to-\infty)=f_0^c(X,a)$, $\partial_Yf_1^*(X,Y\to+\infty)=\partial_yf_0^p(X,a)$, etc \cite{ward1993strong}. 

We now describe the solution in $d=2$ dimensions, all results will be generalized to $d=3$ in the SM \cite{supmat}. 
For the flow, at leading order in the central region, it is readily identified to be the Poiseuille flow induced by the pressure gradient with effective vanishing velocities at $y=a$:
\begin{align}
&\ve[u]_0= U\left(1-\frac{y^2}{a^2}\right)\ve[e]_x, &U=-\frac{a^2(\nabla \Pi)_\infty}{2\eta}.
\end{align} 
Then, in the inner region, the flow satisfies 
\begin{align}
&\tilde{\nabla}\cdot\ve[u]^*=0, & \eta\tilde{\nabla}^2 \ve[u]^*-\tilde{\nabla}\Pi^*   =0,
\end{align}
where $\tilde{\nabla}=\ve[e]_x\partial_X+\ve[e]_y\partial_Y$ and $\Pi=L\Pi^*$ in the inner region. Furthermore, $\ve[u]^*(X,Y)$ vanishes for $Y>0$ and $X=-1/2+n$, with $n$ integers, and far from the inner region one must recover the flow in the central region, so that $\ve[u]^*(X,Y\to-\infty)\simeq-\frac{2U }{a} Y   \ve[e]_x$. The problem for $\ve[u]^*$ is therefore formally equivalent to that of a two dimensional flow next to a periodic array of parallel semi-infinite lines, with  a constant shear far from the plates. This problem has been solved independently in papers by Luchini \cite{luchini1991resistance} and Jeong \cite{jeong2001slip} in which the explicit form of $ \ve[u]^*$ can be found.  Note that, for $Y\to\infty$, the flow takes the form of a series of eddies whose magnitude decreases exponentially with $Y$, which means that we can safely consider that the flow vanishes at leading order in the lateral pores, as anticipated in \cref{eq:ExpFlow}. These eddies can actually be observed on Fig.\ref{fig:Fig1}(b).
For $Y\to-\infty$, the flow becomes
\begin{align}
\ve[u]^*(X,Y\to-\infty)=\frac{2U}{a} \left(-Y+ \frac{\beta}{2} \right) \ve[e]_x,
\end{align} 
with $\beta\simeq0.1772$, which means that there is an effective slip velocity $u_s=UL\beta/a$ at the walls; this phenomenon is well known as the  reduction in drag by surface roughness, and the flow at first order in the central region is $\ve[u]=\ve[u]_0+u_s\ve[e]_x $. This enables one to define an effective slip length 
\begin{align}
b=\frac{u_s}{\vert\partial_y u_x\vert_{y=a}}= \frac{ \beta L}{2},
\end{align}
which is an intrinsic property of the local surface geometry since it neither depends on the flow nor on the width of the channel. We now apply the same approach to find   parameters describing effective boundary conditions for the transport. 

Next, the equations for $f_i$ in the central and the peripheral regions are recurrence equations, here written for $n=0,1,2,3$: 
\begin{align}
&D(\partial_X^2 f_n^w+\partial_y^2 f_{n-2}^w)\nonumber \\& - \sum_{m=0}^{n-1} u_{x,m} \partial_X f_{n-1-m}^w=v_{n-2}-u_{x,n-2},\\
&[(\partial_X h) (\partial_X f_n^p-\delta_{n,1})-\partial_y f_{n-2}^p]_{y=h(X)}=0 
\end{align}
(all terms with negative indices are  by convention zero), with $w\in\{c,p\}$ and $v=\sum_{n\ge0}v_n L^n$. In the inner region, $f^*$ satisfies the Laplace problem
\begin{align}
&(\partial_X^2+\partial_Y^2)f_n^* =0, 
&(\partial_Xf_n^*)_{X=\pm 1/2;Y>0}=\delta_{n,1}.  \label{eq:EqfStar}
\end{align}
In Appendix \ref{sec:CalculDe}, we find the solution of these equations, for successive values of $n$, via complex analysis for \cref{eq:EqfStar} (generalizing the analysis of \cite{mangeat2017geometry}, without flow, to our situation where the matching conditions are modified by the flow). The diffusivity is finally given by 
\begin{align}
D_e&=\frac{ D a +D L (\ln2)/\pi}{a+\delta} \nonumber\\
&+ \frac{4U^2a^2}{9D(a+\delta)^3}    \Bigg\{
\frac{17 a \delta^2}{35} +\frac{6 a^2\delta}{35}  +   \frac{ 2 a^3 }{105}    +  \tau D \delta      \Bigg\}\nonumber\\ 
&+ \frac{4 a^2   U u_s }{45D(a+\delta)^3} \Bigg\{6a\delta^2 +    a^2 \delta   + 15  D \tau \delta   \Bigg\}, \label{eq:DeAfterMap}
\end{align}
with 
\begin{equation}
\delta=\langle h-a \rangle,\, \tau= \int_a^{h_m} \frac{dy}{\delta D W(y)} \bigg[\int_y^{h_m} dy' W(y') \bigg]^2, \label{eq:Def_tau}
\end{equation}
where $\langle g(X) \rangle=\int_{-1/2}^{1/2}dX g(X) $ is the uniform average over the period and $W(y)$ is the width of the lateral pore at distance $y>a$ from the channel center (divided by $L$).

\section{Correspondence with uniform channels}
\label{SecUniform}
\subsection{Uniform channel with sticky boundaries} 
\textcolor{black}{In order to obtain a physical interpretation of the parameters $\tau$ and $\delta$, we}
 now consider a model in which the particle diffuses between two \textit{flat} boundaries which are sticky, with attachment and detachment rates $k_a$ and $k_d$ respectively. In the bulk of the channel, particles diffuse with diffusion coefficient $D$ in a velocity field $u_e(y)\ve[e]_x$, while on the surface they diffuse with  surface diffusion coefficient $D_s$. In this model, the bulk and surface probability density functions $p_b$ and $p_s$ obey
\begin{align}
&\partial_t p_b=-u_e(y)\partial_xp_b  + D (\partial_x^2+\partial_y^2)p_b& (\vert y\vert <a),\\
&\partial_t p_s=D_s\partial_x^2 p_s - k_d p_s+k_a p_b,& (\vert y\vert =a) 
\end{align}
and the boundary condition $D \ve[n]\cdot\nabla p_b= k_a p_b-k_d p_s$ at $y=\pm a$ ensures that the total probability is conserved. Dispersion in such kind of models has been widely studied~\cite{levesque2012taylor,venditti2022exact,jiang2022analytical,hlushkou2014effect,berezhkovskii2013aris}. 
Using the formulas of Ref.~\cite{levesque2012taylor}, we  show in Appendix \ref{sec:FlatSticky} that $D_e$ for this model with flat sticky walls is exactly \cref{eq:DeAfterMap}, as soon as we identify the following parameters
\begin{align}
&u_e=u_x^{(0)}(y)+u_s,\, &D_s= \frac{D L \ln2}{ \pi\delta}, \\
& \delta=\frac{k_a}{k_d},\, &\tau=k_d^{-1}.\label{Param1} 
\end{align}

The length $\delta$ can be identified as the adsorption length for the effective model by the following simple argument. 
In the model with flat sticky walls, the fraction of time that a particle spends in the bound state is clearly $\mu=\delta/(a+\delta)$, while in the full problem the fraction of time spent in the lateral regions is $\mu=\langle h-a\rangle/\langle h\rangle$, because the steady state probability density function is uniform due to the flow incompressibility. Comparing these two expressions for $\mu$ leads to $\delta=\langle h-a\rangle$. 
Note that this length $\delta$ is the area of the pores divided by $L$, so that it does not depend on the dimension $a$ of the open part of the channel. 
Next, the expression of $\tau$ can be rationalized by noting that $\tau$ is the average first escape time out of a lateral pore, starting with equilibrium initial condition in the pore. Indeed, when it is in a lateral pore,  for small $L$, the motion of the particle is effectively a one-dimensional diffusion in an effective entropic potential  $\varphi(y)=-k_\text{B}T \ln W(y)$, with $k_\text{B}$ the Boltzmann’s constant and $T$
 the temperature. Then the mean first escape time $\tau$ to escape at $y=a$ starting from an equilibrium distribution in the lateral pore is known and is exactly \cref{eq:Def_tau}, see Eq.~(2.20) in  Ref.~\cite{szabo1980first}. Finally, the effective surface diffusivity $D_s$ in \cref{Param1} takes into account the fact that particles in the vicinity of the line $y=a$ are not completely immobile. 

\subsection{Uniform channel with anisotropic diffusion} 
\label{sectionAnisotropic}
\textcolor{black}{Alternatively, all terms of the effective diffusivity (\ref{eq:Def_tau}) can also be obtained by considering a particle diffusing in a uniform channel with anisotropic diffusivity. More specifically, we can approximate the motion in the longitudinal direction $x(t)$ in  the real channel as satisfying the stochastic differential equation
\begin{equation}
\dd x_t =  \sqrt{2 D_\parallel(y_t)} \,\dd B_{\parallel,t} +  u_e(y_t) \theta(a-\vert y\vert)\, \dd t,
\end{equation}
where $y_t$ is the distance to the central axis and $\langle \dd B_{\parallel,t}^2\rangle=\dd t$. Here, we have 
\begin{equation}
D_\parallel(y) = D \theta(a+\ell-\vert y\vert),
\end{equation}
where $\theta$ is the Heaviside step function. This means that the particle only diffuses when it is in the channel part $\vert y\vert <a+\ell$ and we ignore its dispersion in the interior of the lobe or corrugated part, which is a valid approximation for the form of strong corrugation used here. The length $\ell$ is a {\em diffusive incursion length}  representing the distance from the protrusion entrance under which the particle, while it is still  inside the protrusions, can nevertheless be considered as undergoing free diffusion along the channel. Similarly, the drift vanishes when $\vert y\vert>a$. The  process in the lateral direction $y_t$ obeys
\begin{align}
&\dd y_t = \sqrt{2D} \,\dd B_{\perp,t} + \frac{D}{k_\text{B} T}\varphi'(y_t) \, \dd t, \\
&\varphi(y)=
\begin{cases}
-k_\text{B}T\ln [(y/a)^{d-2} W(y)]&(\vert y\vert >a),\\
-k_\text{B}T\ln [(y/a)^{d-2} ]&(\vert y\vert <a),
\end{cases}
\end{align}
where $\langle \dd B_{\perp,t}^2\rangle=\dd t$, $W(y)L$ is the longitudinal distance between the pore's wall at distance $y$ from the center (divided by $L$). The above equation is valid for $d=2$ or $d=3$ and takes into account the fact that the number of configurations at fixed $y$ is $\omega(y)= y^{d-2} W(y)$ (up to a multiplicative factor), giving rise to an entropic potential $\varphi(y)=-k_\text{B}T\ln\omega=-k_\text{B}T\ln y^{d-2} W(y)$ which couples a Fick-Jacobs approximation for the motion in the lateral direction, and the fact that $y$ is the radial part of a Brownian motion. The above equation is derived for $\vert y\vert >a$ and is trivially extended for $\vert y\vert<a$ where the part due to $W(y)$ is irrelevant and diffusion is free.  }

\textcolor{black}{This approximate problem is therefore just an effective Taylor dispersion problem in a channel of height $h_{m}=\max(h)$ with spatially dependent anisotropic diffusion tensor. The effective diffusivity for this problem is given in Ref. \cite{alexandre2021generalized} as
\begin{align}
D_e=& \int_0^{h_m}\dd y D_\parallel(y)p_e(y)\nonumber\\
+&\int_0^{h_m}\dd y \frac{\left\{\int_y^{h_m}\dd y_1 p_e(y_1)[u_e(y_1)-v]\right\}^2}{D_\perp(y)p_e(y)} \label{eq:DeMappingNonUniformDiffChann},
\end{align}
where   $p_e(y)$  is the stationary distribution of $\vert y\vert$,
\begin{align}
&p_e(y)=\frac{e^{-\varphi(y)/k_\text{B}T}}{\int_0^{h_m} \dd y_1 e^{- \varphi(y_1)/k_\text{B}T}},  \label{Def_pe}
\end{align}
and $v=\int dy p_e(y) u_e(y)$ the average longitudinal velocity. Note that here we consider the line $y=0$ as reflecting for the particle, this does not change the value of the final diffusivity. Applying these formulas for $u_e=[U(1-y^2/a^2)+u_s]\theta(a-\vert y\vert)$ and $d=2$, we show in Appendix \ref{sec:AnisotropicFlat} that the diffusivity (\ref{eq:DeMappingNonUniformDiffChann}) corresponds exactly to Eq.~(\ref{eq:DeAfterMap}) at order $L$ if one identifies the length $\ell$ as
\begin{equation}
\ell=L\ln 2/\pi,
\end{equation}
which takes a universal value. The value of $D_s$ in our effective sticky wall description can be recovered as being $D_s/D=\ell/\delta$, i.e. the fraction of volume protrusion where the particle is {\em freely} diffusing. The concept of diffusive incursion length thus generalizes to particle's diffusion the concept of slip length in hydrodynamics.}
 
\section{Numerical validation} 
\label{SecNum}
\begin{figure}
\includegraphics[width=8cm]{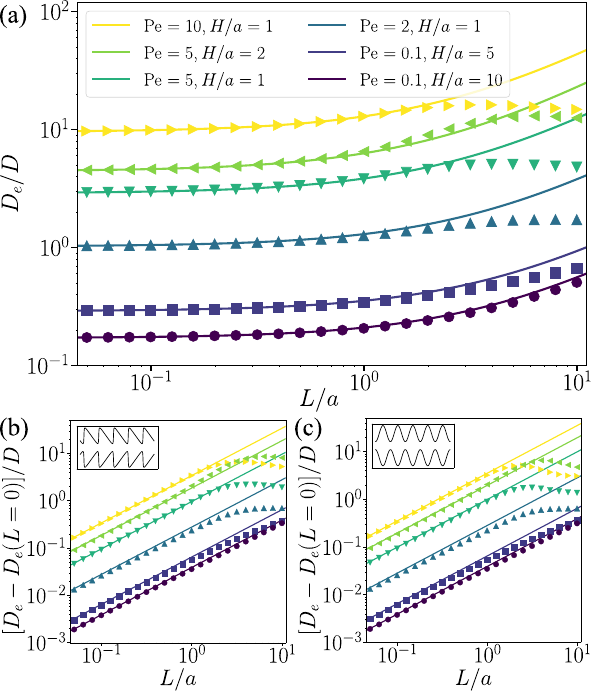}
\caption{(a) Effective diffusivity as a function of the period for a two-dimensional channel of profile 
$h(X)=a + H/2\left(\arctan\left\{\cos(2\pi X) / \left[\sin(2\pi X)+1.1\right]\right\}/1.1411+1\right)$, shown in the inset of (b).
Symbols: numerical evaluation of the exact equations; lines: theoretical prediction \cref{eq:DeAfterMap}. The values of the P\'{e}clet number $\text{Pe}=2Ua/3D$ and the corrugation depth $H/a$ are given in the legend. (b) Value of $D_e$ after substraction of its theoretical value at $L=0$ for the same channel as in (a). Symbols: numerics; lines: theoretical first order prediction. (c) Same quantity as (b) for a sinusoidal channel with profile $h(X)=a+H[1+\cos(2\pi X)]/2$, for the same parameters.}
\label{fig:Fig2}
\end{figure}

In Fig.~\ref{fig:Fig2}(a), we see an excellent agreement between the  numerically obtained $D_e$ (solving Eqs. (\ref{eq:ExprDe}), (\ref{eq:EqForf}) and (\ref{eq:BCForf}) via a finite element PDE solver) and the analytical result (\ref{eq:DeAfterMap}) for various  corrugation depths $H$ and flow magnitudes $U$,  
in the limit of small $L$. Remarkably, the predicted value holds for not-too-small values of $L$ - for most curves the agreement is good up to $L\simeq 2a$, when the period is comparable to the diameter of the channel. 
The validity of the results here at order $L$ is verified in Figs.~\ref{fig:Fig2}(b) and (c), where we show for two different channel profiles that $D_e-D_e(L=0)$ depends linearly on $L$ for small $L$, with prefactors correctly predicted by  Eq. (\ref{eq:DeAfterMap}).

\section{Generalization to 3D axisymmetric channels}
\label{Gen3D}
\textcolor{black}{Our approach can be generalized to the case of three-dimensional axisymetric channels, calculation details can be found in Appendix \ref{sec:3DTheory}. We find that, in the limit of strongly corrugated channels, the flow  in the central region it is given by
\begin{align}
&\ve[u]=U(1-y^2/a^2+\beta L /a) \ {\ve[e]}_x + \mathcal{O}(L^2),
\end{align}
where $U=-a^2 (\nabla \Pi)_\infty/(2\eta)$. The diffusivity in the strong corrugation limit reads
\begin{align}
D_e=&\frac{ D a}{a+2\delta}+  \frac{DL 2\ln 2}{\pi \, (a+2\delta) }\nonumber\\
&+\frac{U^2a^2\left(
 a^3+12a^2\delta+44a\delta^2 + 96  D\tau \delta \right)}{192 D\,(a+2\delta)^3}    \nonumber\\
& +\beta L a U^2 \frac{a^2 \delta + 8 a \delta^2 + 24 D\delta\tau  }{12D \,(a + 2 \delta)^3}\label{eq:De3D},
\end{align}
where
\begin{align}
&\langle h^2\rangle=a^2+ 2 a\delta,\\
&\tau=\frac{1}{D\delta a}\int_a^{h_m}dy\frac{\left(\int_y^{h_m}dy' W(y')y' \right)^2}{W(y)y},
\end{align}
Equation (\ref{eq:De3D}) corresponds to the result obtained for flat sticky walls in Ref.~\cite{levesque2012taylor} (when one neglects the slip velocity), when one identifies $\delta=k_a/k_d$; $\tau=1/k_d$, and the effective surface diffusivity $D_s=\frac{D L \ln2}{ \pi\delta}$. Note that $\tau$ is again interpreted as the escape time out of a lateral pore. 
Alternatively, one can also compare with the model of uniform channels of Section \ref{sectionAnisotropic}, this leads to the diffusive incursion length $\ell=L\ln 2/\pi$, exactly as in the 2D case, just as the effective slip length $b=\beta L/2$ for the flow is the same for 2D and 3D channels.
}

\section{Conclusion}
 We have studied the dispersion of particles in a channel in the strongly corrugated limit (small period limit) in the presence of a pressure driven flow. In this limit, all standard approximations (Fick-Jacobs, lubrication) fail, and we have found an explicit formula for effective transport coefficients which is exact at next-to-leading order in $L$. At this order, we find that transport here is equivalent to  transport in a perfeclty flat but partially sticky tube. The effective attachement and detachment parameters are explicitly determined in terms of the geometry of the lateral regions, and of the exit time out of the lateral pores. Furthermore, the analysis at next-to-leading order in $L$ enables us to see  that (i) the effective flow is non-zero in the vicinity of the entrance to the lateral pore, and (ii) that when in a lateral branch the dispersion  of particles along  the longitudinal direction is not completely suppressed due to  the finite size of the pore entrance. The first  point (i) leads to the emergence of an effective slip length, while the second (ii) leads to an effective surface diffusivity that is determined by a universal numerical constant, or equivalently to the definition of the length at which the particles at the entrance of protrusions can nevertheless be considered as mobile.

\begin{acknowledgments}
T. G. acknowledges support of the grant \textit{ComplexEncounters}, ANR-21-CE30-0020. D. S. D. acknowledges support of the grant \textit{EDIPS}, ANR-23-CE30-0020.
\end{acknowledgments}

\textcolor{black}{The data that support the findings of this article are openly available \cite{Alexandre2025DATA}.} 
 
 \appendix
 
\section{Identification of the stochastic equation with surface terms}
\label{sec:SecFormalism}

Here, we briefly explain how to show that the effective drift, taking into account the influence of boundaries, is 
\begin{align}
 \ve[V]=\ve[u]+D\, \ve[n]\, \delta_s(\ve[r]), \label{eq:DefV}
\end{align}
meaning that the stochastic differential equation (SDE) for the position $\ve[r]_t$ of the particle at time $t$  can be written as
\begin{align}
&\dd\ve[r]_t= \ve[V](\ve[r]_t) \,\dd t +\sqrt{2D}\,\dd B_t, &  \overline{\dd B_t} =0, \   \overline{\dd B_t^2}=\dd t. 
\end{align}
Let us identify the generator $G$, defined as the operator such that the evolution of any test function $\phi(\ve[r])$ reads
\begin{align}
\partial_t \overline{\phi}= \overline{G \phi},
\end{align}
and it is related to the SDE by 
\begin{align}
G=\ve[V]\cdot\nabla+D\nabla^2  \label{eq:DefG}.
\end{align}
Using the Fokker-Planck equation (\ref{FkPEq}), we find 
\begin{align}
\partial_t  \overline{\phi} &=\partial_t \int_\Omega \dd\ve[r]\, P(\ve[r],t)\phi(\ve[r])\nonumber\\
&= \int_\Omega \dd\ve[r] \, \phi(\ve[r])\left\{-\nabla \cdot[\ve[u]\, P -D\nabla P]\right\}.
\end{align}
Using integration by parts (divergence theorem) and the boundary conditions for $P$, we obtain
\begin{align}
\partial_t\overline{ \phi}= \int_\Omega \dd\ve[r] \left\{  P(\ve[r],t)  (\ve[u]\cdot\nabla +   D\nabla^2)\phi(\ve[r]) \right\}\nonumber\\
+ \int_{\partial\Omega} \dd S(\ve[r])\ve[n]\cdot D P(\ve[r],t)  \nabla\phi .
\end{align}
We can see the boundary term as a bulk term, so that we can identify the generator $G$ as
\begin{align}
\partial_t\overline{ \phi}&=  \int_\Omega \dd\ve[r]   P(\ve[r],t)\left\{  [\ve[u]+D\,\ve[n]\,\delta_s(\ve[r])]\cdot\nabla +   D\nabla^2\right\}\phi(\ve[r]) \nonumber\\
&=\overline{G\phi},
\end{align}
so that $G$ is given by \cref{eq:DefG}, with an effective drift field $\ve[V](\ve[r])$  given by \cref{eq:DefV}. 

\section{Calculation of the effective diffusivity in the strong corrugation limit}
\label{sec:CalculDe}
\subsection{Expression of $f$ in the peripheral and central regions}

Here we describe how to compute the auxiliary function $f$. First, in the peripheral and the central regions, the relevant variable in the longitudinal direction is $X=x/L$, while $y$ is considered to be of order $a$. In terms of these variables, the equations become
\begin{align}
-\frac{u_x}{L} \partial_X f  + D\left(\frac{1}{L^2} \partial_X^2 + \partial_y^2 \right)f= v-u_x,  \label{eq:EqfRescaled}
\end{align} 
in both peripheral and central regions, with $v=\overline{u}_x$. In the peripheral region, the boundary condition is
\begin{align}
& \frac{1}{L}(\partial_X h) \left(\frac{\partial_Xf^p}{L}-1\right)-\partial_yf^p =0 & [y=h(X)] \label{eq:BCRescaled}.
\end{align}
while for the central region one has the condition that $f^c$ is periodic of period $1$ and the condition $\partial_yf^c=0$ at $y=0$ (imposed by symmetry). Note that here $h$ is considered as a function of $X=x/L$.   

Now, inserting the series expansion in powers of $L$, $f^w=\sum_{n\ge0} L^n f_n^w(X,y)$ ($w\in \left\{c,p\right\}$) in both central and peripheral regions into \cref{eq:EqfRescaled,eq:BCRescaled}, we find  
\begin{align} 
&D\partial_X^2 f_0^c=0, 						&D\partial_X^2 f_0^p=0,\label{eq:Eqf0}\\
&D\partial_X^2 f_1^c-u_x^{(0)} \partial_X f_0^c =0,  &D\partial_X^2 f_1^p =0,  \label{eq:Eqf1}\\
&f_0^c(X+1,y)=f_0^c(X), & (\partial_Xf_0^p)_{y=h(X)}=0\label{eq:BC0},\\
&f_1^c(X+1,y)=f_1^c(X), &(\partial_Xf_1^p)_{y=h(X)}=1. \label{eq:BC1}
\end{align}
These equations lead to
\begin{align} 
&f_0^{c}(X,y)=f_0^{c}(y) , & f_0^{p}(X,y)=f_0^{p}(y), \\
&f_1^{c}(X,y)=f_1^{c}(y) , & f_1^{p}(X,y)=X+b_1^p(y),
\end{align}
where $f_0^c,f_0^p,f_1^c,b_1^p$ are functions of $y$ only. To determine these functions, we need to consider the next orders in the power expansion of \cref{eq:EqfRescaled,eq:BCRescaled}. First, for the peripheral region (where $u_x=0$), we obtain
\begin{align}
&D[\partial_X^2 f_2^p+\partial_y^2f_0^p]  =v_0, \\
&D[\partial_X^2 f_3^p+\partial_y^2f_1^p] = v_1.
\end{align}
Integrating over the variable $X$, we obtain
\begin{align}
&D\partial_X f_2^p =[v_0-D(f_0^p)''(y)]X+A_2^p(y), \label{eq:Eqdxf2}\\
&D\partial_X f_3^p  =[v_1-D(b_1^p)''(y)]X+A_3^p(y),\label{eq:Eqdxf3} 
\end{align}
where $A_2^p$ and $A_3^p$ are functions of $y$ only.  The boundary conditions are
\begin{align}
&h'(X)\partial_X f_2^p-\partial_yf_0^p=0 & [y=h(X)],\\
&h'(X)\partial_X f_3^p-\partial_yf_1^p=0& [y=h(X)],
\end{align}
so that, using \cref{eq:Eqdxf2,eq:Eqdxf3}, we obtain
\begin{align}
h'(X)\{[v_0-D(f_0^p)''(h(X))]&X+A_2^p (h(X)) \}\nonumber\\
&= D (f_0^p)'(h(X)), \\
h'(X)\{[v_1-D(b_1^p)''(h(X))]&X+A_3^p (h(X))\} \nonumber\\
&= D (b_1^p)'(h(X)). 
\end{align}
We remark that we can integrate these equations once to obtain 
\begin{align}
& D X(f_0^p)'(h(X)) =v_0 \int_{-1/2}^X \dd w h'(w)  w + G_2^p(h(X))  \label{eq:05492}, \\
& D X(b_1^p)'(h(X)) =v_1 \int_{-1/2}^X \dd w h'(w)  w + G_3^p(h(X)),
\end{align}
where $G_2^p$ and $G_3^p$ are primitive functions of $A_2^p$ and $A_3^p$, respectively: $\partial_y[G_i^p(y)]=A_i^p(y)$ ($i\in\left\{2,3\right\}$) and $G_2^p$ and $G_3^p$ are determined up to an additive constant (unknown so far).

\begin{figure}
    \centering\includegraphics[width=0.8\linewidth]{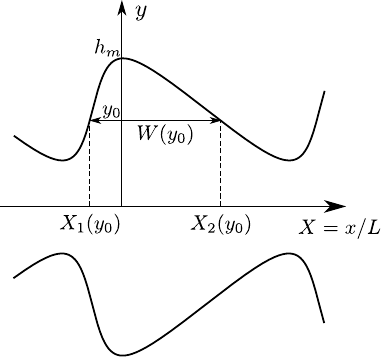}
    \caption{Illustration of the different quantities $X_1$, $X_2$ and the width $W$.}
    \label{fig:width_W}
\end{figure}
Now, for a given value of $y$, we can find two values of $X$, say $X_1(y)$ and $X_2(y)$ so that $y=h(X_1(y))=h(X_2(y))$, with $X_1(y)<X_2(y)$, see \cref{fig:width_W}. 
If we write the above equations for $X_1(y)$ and $X_2(y)$ and take the difference between the two, we obtain
\begin{align}
& D[X_2(y)-X_1(y)]  (f_0^p)'(y) =v_0  \int_{X_1(y)}^{X_2(y)}\dd w \, w\, h'(w), \label{eq:Res_f0p} \\
& D [X_2(y)-X_1(y)] (b_1^p)'(y) =v_1 \int_{X_1(y)}^{X_2(y)}\dd w \, w\, h'(w).\label{eq:Res_b1p}
\end{align}
Using the change of variable $y_1=h(w)$ in the above integrals [after having separated the integration interval between the intervals $[X_1(y);X_1(h_m)]$ and $[X_2(h_m);X_2(y)]$, where $h_m=\text{max}(h)$], we remark that
\begin{align}
\int_{X_1(y)}^{X_2(y)}\dd w \, w \,h'(w) =- \int_y^{h_m}\dd y_1 W(y_1),
\end{align}
where $W(y)=X_2(y)-X_1(y)$ is the local width of the lateral pore (divided by $L$), see \cref{fig:width_W}. We thus obtain 
 \begin{align}
& D  (f_0^p)'(y) =- \frac{v_0}{W(y)}\int_y^{h_m}\dd y_1 W(y_1)  \label{eq:Res_f0p}, \\
& D  (b_1^p)'(y) =- \frac{v_1}{W(y)}\int_y^{h_m}\dd y_1 W(y_1) \label{eq:Res_b1p}.
\end{align}

For the central region, we have 
\begin{align}
&D[\partial_X^2 f_2^c+\partial_y^2f_0^c(y)]  =v_0-u_x^{(0)}(y),\\
&D[\partial_X^2 f_3^c+\partial_y^2f_1^c(y)]  =v_1-u_x^{(1)}(y).
\end{align}
The fact that $f_2^c$ and $f_3^c$ are periodic functions of $X$ imposes
\begin{align}
D(f_0^c)''(y)=v_0-u_x^{(0)}(y),\\
D(f_1^c)''(y)=v_1-u_x^{(1)}(y).
\end{align}
Therefore, noting that $\partial_y f =0$ at $y=0$ (by symmetry, at all orders), we have
\begin{align}
&D\partial_y f_0^c (y)=v_0 y - U\left(y-\frac{y^3}{3a^2}\right)\label{eq:Res_f0c},\\
&D\partial_y f_1^c (y)=\left(v_1-\frac{U\beta}{a}\right)y \label{eq:Res_f1c}.
\end{align}
Here it is useful to note that
\begin{align}
&v_0= \frac{2\, U\,a  }{3\,\langle h\rangle},  & v_1=\frac{U\, \beta}{\langle h\rangle},
\end{align}
and that the area of the lateral region can be written  as
\begin{align}
\langle h\rangle -a=\frac{1}{W(a)}\int_a^{h_m}\dd y' W(y') = \int_a^{h_m}\dd y' W(y'),
\end{align}
with $W(a) = 1$. Using these values, we see  by comparing \cref{eq:Res_f0c,eq:Res_f0p} that $\partial_y [f_0^c-f_0^p]_{y=a}=0$, so there is no discontinuity in the derivative of $f_0$ at $y=a$ at this order. The same property holds for $f_1$. We can thus impose that $f_0$ is regular (continuous; with continuous derivative) at $y=a$, this leads to 
\begin{align}
&Df_0(y)=\nonumber\\
&\begin{cases}
Df_0^c(y)=(v_0-U) \frac{y^2-a^2}{2} + U \frac{(y^4-a^4)}{12 a^2}+C_0,& (y<a),\\
Df_0^p(y)=- \int_a^y \frac{d y' v_0}{W(y')} \int_{y'}^{h_m} \dd y'' W(y'') +  C_0, & (y>a),
\end{cases}\label{eq:Resultf0}
\end{align}
where the integration constant $C_0$ is fixed by the normalization condition $\int_\Omega \dd\ve[r] f =0$, which leads to
\begin{align}
\int_0^a \dd y f_0^c(y)+\int_a^{h_m}\dd y W(y)   f_0^p(y)=0.
\end{align}
Using the previous result for $f_0^c,f_0^p$, the above equation  leads to
\begin{align}
C_0 & =\frac{(-1)}{\langle h\rangle } \Bigg\{\frac{4U a^3}{15}-\frac{a^3 v_0}{3}\nonumber\\
& - v_0 \int_a^{h_m}\dd y\frac{\left[\int_y^{h_m}\dd y''W(y'')\right]^2}{W(y)}  \Bigg\}. \label{eq:ValueC0}
\end{align}

\subsection{Expression of $f$ in the inner region}

At this stage, we have completely determined $f_0$, but the expressions for $f_1(X,y)$ in the peripheral and central regions cannot be matched, because $f_1^p$ depends on $X$ whereas $f_1^c$ does not. This suggests that we look for solutions in the inner layer of the form
\begin{align}
f(x,y)\simeq f_0(a)+L f^*(X,Y).
\end{align}
This form is imposed by the matching conditions with the central and peripheral regions:
\begin{align}
 f^*(X,Y)\underset{Y\to+\infty}{\simeq}& f_1^p(X,a) + Y f_0'(a)\nonumber\\
 &=[X+b_1^p(a)] + Y f_0'(a), \label{BCfStar1}\\
  f^*(X,Y)\underset{Y\to-\infty}{\simeq}& f_1^c(a) + Y f_0'(a). \label{BCfStar2}
\end{align}
The equation for $f^*$ is found by expanding \cref{eq:EqForf} in powers of $L$ (after having rescaled $x,y$ by $L$, using $X=x/L,Y=(y-a)/L$), this leads to
\begin{align}
(\partial_X^2+\partial_Y^2)f^* =0.
\end{align}
Furthermore, $f^*$ must be periodic in $X$ (with period $1$), and the boundary conditions are 
\begin{align}
(\partial_Xf^*)_{X=\pm 1/2;Y>0}=1,
\end{align}
where we have taken the position of the values at which $h$ is minimal at $X=1/2+n$. 
In the absence of the term $Y f_0'(a)$ in the matching conditions (\ref{BCfStar1}) and (\ref{BCfStar2}), this problem was solved in the context of dispersion without flow in \cite{mangeat2017geometry}, notably it also arises in the context of the calculation of the drag reduction when the direction of the flow is parallel to semi-infinite plates \cite{bechert1989viscous}. 
Taking into account the term $Y f_0'(a)$ slightly modifies the solution for $f^*$, which is
\begin{align}
f^*(X,Y)&=Y f_0'(a) + f_1^c(a) \nonumber\\
&+\text{Re}\left[\frac{\text{i}}{\pi}\ln\left(1+\sqrt{1+e^{-2\pi \text{i} (X+\text{i} Y)}}\right)\right] ,  \label{eq:fStar}
\end{align}
where $\text{Re}(z)$ is the real part of a complex number $z$, and $\text{i}^2=-1$.  Note that  this solution can be matched with the peripheral solution under the condition
\begin{align}
b_1^p(a)=f_1^c(a),
\end{align}
so that $f_1^p$ is equal to $f_1^c$ at the position in the middle position between the necks of the channel ($X=0$). With this matching condition, we  obtain 
\begin{align}
&D f_1^c (y)=\left(v_1-\frac{U\beta}{a}\right)\frac{y^2-a^2}{2}+C_1, \label{eq:Resultf1c}\\
&D b_1^p(y)=-v_1 \int_a^y \frac{\dd y'}{W(y')}\int_{y'}^{h_m}\dd y'' W(y'')  + C_1.
\end{align}
The constant $C_1$ is found by requiring that $\int_V f \dd\ve[r]=0$ at order $L$, so that
\begin{align}
\int_0^a \dd yD f_1^c(y)+\int_{-1/2}^{1/2} \dd X \int_a^{h(X)} \dd y\  [Db_1^p(y)+DX]=0.
\end{align}
This leads to
\begin{align}
C_1 =&-\frac{1}{\langle h\rangle } \Bigg\{    -\left(v_1-\frac{U\beta}{a}\right)  \frac{a^3}{3}  +D\langle X h(X) \rangle \nonumber\\
& -v_1 \int_a^{h_m} \frac{\dd y'}{W(y')}\left[\int_{y'}^{h_m}\dd y'' W(y'') \right]^2 \Bigg\}.
\end{align}
 At this stage, we have fully determined $f_0$ and $f_1$ in all the regions of the channel. To compute the diffusivity at next-to-leading order in $L$, we also need to compute $\partial_X f_2$ at the channel boundary, which can be obtained from \cref{eq:Eqdxf2}:
\begin{align}
D[\partial_X f_2^p]_{y=h(X)}  =[v_0-D(f_0^p)''(y)]X+\partial_y G_2^p(y).
\end{align}

\subsection{Expression of the diffusivity}

The effective diffusivity $D_e$ is computed as follows:
\begin{align}
D_e&=D+ \frac{1}{\Omega}\int_\Omega \dd\ve[r] \,  [u_x f-D\partial_x f]\nonumber\\
&=D+ \int_{-1/2}^{1/2}\frac{\dd X}{\langle h\rangle} \int_0^{h(X)} \dd y \,  \left[u_x f-\frac{D}{L}\partial_X f\right].\label{eq:DiffGen}
\end{align}
Let us write the general expansion
\begin{align}
D_e=D_e^{(0)}+L [D_A+D_B+D_C]+\mathcal{O}(L^2),
\end{align}
where the leading-order term reads:
\begin{align}
D_e^{(0)}=D+ \int_{-1/2}^{1/2}\frac{\dd X}{\langle h \rangle}\left[ \int_0^{a} \dd y\,  u_x^{(0)} f_0^c   -D \int_{a}^{h(X)}\dd y \, \partial_X f_1^p   \right],
\end{align}
where this expression takes into account the fact that $u_x$ vanishes in the peripheral region, while $\partial_X f_0^p=\partial_X f_0^c=0$, and $\partial_X f_1^c$ vanishes in the central region.  The next-to-leading order components to the effective diffusivity read:
\begin{align}
&D_A= \frac{1}{\langle h\rangle}\int_{-1/2}^{1/2}\dd X    \int_0^{a} \dd y\, [u_x^{(0)}f_1^c +u_x^{(1)}f_0^c ],  \\
&D_B=-\frac{D}{\langle h \rangle} \int_{-1/2}^{1/2}\dd X     \int_{a}^{h(X)}\dd y \, \partial_X f_2^p, \label{eq:DBTerm}\\
&D_C=-\frac{D}{\langle h \rangle} \int_{-1/2}^{1/2}\dd X       \int_{-\infty}^\infty \dd Y\, [ \partial_X f^*(X,Y)-\theta(Y) ], \label{eq:DCTerm}
 \end{align}
 where $\theta(\cdot\cdot\cdot)$ is the Heaviside theta function. 
Note that the last integral is the contribution due to the inner layer $f^*$. To make explicit the origin of $D_B$ and $D_C$, we consider  the integral
\begin{align}
J=\int_{-1/2}^{1/2}\int_0^{h(X)}\dd y \,\partial_X f(X,y),
\end{align}
which is one of the components of the effective diffusivity in \cref{eq:DiffGen}. At leading order in $L$, we have $J\simeq J_0$ with
\begin{align}
J_0=&\int_{-1/2}^{1/2}\int_0^{a}\dd y \,\partial_X f_1^c(X,y)\nonumber\\
&+\int_{-1/2}^{1/2}\int_a^{h(X)}\dd y \,\partial_X f_1^p(X,y).
\end{align}
To calculate the next-to-leading order, we define the intermediate length $\varepsilon$ with $L\ll\varepsilon\ll a$, and we write
\begin{align}
&J-J_0=\nonumber\\
&\int_{-1/2}^{1/2}\dd X\bigg\{\int_0^{a-\varepsilon} \dd y\, \partial_X(f-f_1^c)+\int_{a-\varepsilon}^{0} \dd y\,\partial_X (f-f_1^c) \bigg. \nonumber\\& \bigg.+\int_0^{a+\varepsilon} \dd y\,\partial_X (f-f_1^p)+\int_{a+\varepsilon}^{h(X)} \dd y \,\partial_X(f-f_1^p)\bigg\}.
\end{align}
In each of these regions, we approximate $\partial_X(f-f_1)$ by its expression in the appropriate region of the boundary layer expansion:
\begin{align}
J-J_0&\simeq L \int_{-1/2}^{1/2}\dd X\bigg\{\int_0^{a-\varepsilon} \dd y \,\partial_Xf_2^c \nonumber\\
&+ \int_{-\frac{\varepsilon}{L}}^{0} \dd Y\partial_X [f^*(X,Y)-\partial_Xf_1^p(X,a+YL)]\nonumber\\
&+\int_{0}^{\frac{\varepsilon}{L}} \dd Y\partial_X [f^*(X,Y)-\partial_Xf_1^c(X,a+YL)]\nonumber\\
&+\int_{a+\varepsilon}^{h(X)} \dd y \partial_X f_2^p \bigg\},
\end{align}
where we have used $Y=(y-a)/L$ in the boundary layer. Now, using $\partial_X f_2^c=\partial_X f_1^c=0$,
in the limit  $L\to0$, with $\varepsilon/L\gg1$ and $\varepsilon\ll a$, we obtain
\begin{align}
J-J_0\simeq &L \int_{-1/2}^{1/2}\dd X\Bigg\{ \int_{-\infty}^{0} \dd Y \partial_X f^*(X,Y)\nonumber\\
&+\int_{0}^{\infty} \dd Y\partial_X [f^*(X,Y)-\partial_Xf_1^p(X,a)]\nonumber\\
&+\int_{a}^{h(X)} \dd y \partial_X f_2^p \Bigg\}.
\end{align}
Noting that $\partial_Xf_1^p(X,a)=1$, using the above expression to evaluate the integrals involving $\partial_Xf$ in \cref{eq:DiffGen}, we obtain the terms $D_B$ and $D_C$ defined in \cref{eq:DBTerm,eq:DCTerm}.

Using \cref{eq:Resultf0,eq:ValueC0,eq:Resultf1c}, the integrals in the leading order term can be calculated, leading to
\begin{align}
D_e^{(0)}=\frac{ D a}{\langle h\rangle}+ \frac{4U^2a^2}{9D\langle h\rangle^3}    \Bigg\{
\frac{17 a \langle \eta \rangle^2}{35} +\frac{6 a^2 \langle \eta \rangle}{35}  +   \frac{ 2 a^3 }{105}   \nonumber\\
+  \int_a^{h_m}\dd y\frac{\left(\int_y^{h_m}\dd y' W(y') \right)^2}{W(y)}  \Bigg\},
\end{align}
where we have defined $\eta=h-a$.

Let us now calculate the terms for the diffusivity at next-to-leading order. First, using \cref{eq:Resultf1c}, we obtain
\begin{align}
D_A=&-\frac{v_0\langle X h(X)\rangle}{\langle h\rangle}+\frac{4 a U^2 \beta}{45D \langle h\rangle^3} \Bigg[6a\langle\eta\rangle^2  \nonumber\\
  &+a^2\langle\eta\rangle   + 15 \int_a^{h_m}\dd y\frac{\left(\int_y^{h_m}\dd y' W(y') \right)^2}{W(y)} \Bigg].
\end{align}
Next, using \cref{eq:Eqdxf2} we obtain
\begin{align}
  D_B =  \frac{(-1)}{\langle h \rangle} \int_{-1/2}^{1/2}\dd X     \int_{a}^{h(X)}\dd y \,  \Big\{& [v_0-D(f_0^p)''(y)]X\nonumber\\
  &+\partial_y G_2^p(y) \Big\}.
  \end{align}
Performing the integral over $y$, and using the property $\int_{-1/2}^{1/2}\dd X X=0$, we obtain
\begin{align}
D_B  = -\frac{1}{\langle h \rangle} \int_{-1/2}^{1/2}\dd X     \Big\{& [v_0h(X)-D(f_0^p)'(h(X))]X \nonumber\\
&+ G_2^p(h(X))-G_2^p(a)\Big\}.
  \end{align}
  Using \cref{eq:05492}, we can simplify this expression:
 \begin{align}
  D_B  = -\frac{1}{\langle h \rangle} \int_{-1/2}^{1/2}\dd X    & \Bigg[   v_0 h(X)  X -G_2^p(a)\nonumber\\
  &-v_0 \int_{-1/2}^X \dd w \,w\,h'(w) \Bigg].
 \end{align}
Writing \cref{eq:05492} for $X=-1/2$ and $X=1/2$ leads to
\begin{align}
&G_2^p(a)= -D (f_0^p)'(a)/2 =- \frac{v_0\langle X h'(X)\rangle}{2} .
\end{align}
Using integrations by parts, we also note that
\begin{align}
 \int_{-1/2}^{1/2}\dd X  &        \int_{-1/2}^X \dd w \,w \,h'(w) \nonumber\\
 &=- \int_{-1/2}^{1/2}\dd X X^2h'(X)+\frac{\langle X h'(X)\rangle}{2}\nonumber\\
& = 2\langle X(h(X)-a)\rangle  +\frac{\langle X h'(X)\rangle}{2}.
\end{align}
As a consequence, we obtain for $D_B$ (noting that, trivially, $\langle X\rangle=0$)
 \begin{align}
  D_B  = \frac{v_0}{\langle h \rangle} \langle  h(X) X\rangle. 
 \end{align}
Last, the term $D_C$ can be calculated by performing the integration over $X$ in \cref{eq:DCTerm}:
\begin{align}
D_C=\frac{(-D)}{\langle h\rangle} \int_{-\infty}^\infty \dd Y \left[f^*\left(\frac{1}{2},Y\right)-f^*\left(\frac{-1}{2},Y\right)-\theta(Y)\right].
\end{align}
Using the explicit value of $f^*$ given by \cref{eq:fStar}, we obtain
\begin{align}
D_C&=-\frac{D}{\langle h\rangle} \int_{0}^\infty \dd Y \left[ \frac{2}{\pi}\text{Arctan}\left(\sqrt{e^{2\pi Y}-1}\right) -1 \right]\nonumber\\&=\frac{D \ln 2}{\pi \,\langle h\rangle }.
\end{align}
Interestingly this is the same result as that computed without flow in Ref.~\cite{mangeat2017geometry}, although the function $f^*$ itself is modified by the flow.  

Collecting these results, we find our final expression for the effective diffusivity
\begin{align}
D_e=&\frac{ D a}{\langle h\rangle}+\frac{D L \ln 2}{\pi \,\langle h\rangle }
+\frac{4U^2a^2}{9D\langle h\rangle^3}    \Bigg\{
\frac{17 a \langle \eta \rangle^2}{35} +\frac{6 a^2 \langle \eta \rangle}{35}  \nonumber\\
&+   \frac{ 2 a^3 }{105}   +  \int_a^{h_m}\dd y\frac{\left(\int_y^{h_m}\dd y' W(y') \right)^2}{W(y)}  \Bigg\}
\nonumber \\
&+ \frac{4 a L U^2 \beta}{45D\langle h\rangle^3} \Bigg\{6a\langle\eta\rangle^2+  a^2\langle\eta\rangle    \nonumber\\
&+ 15\int_a^{h_m}\dd y\frac{\left(\int_y^{h_m}\dd y_1 W(y_1) \right)^2}{W(y)} \Bigg\}. \label{eq:De2D}
\end{align}

\section{Effective diffusivity for uniform 2D channels with sticky boundaries}
\label{sec:FlatSticky}

Here we explain how the effective diffusivity for a flat channel with sticky walls can be obtained by using the results of Ref.~\cite{levesque2012taylor}, where in 2D it was obtained that 
\begin{align}
D_e&=  D_b  P_B +D_s P_S\nonumber\\
&+ \int_{-a}^{a} d{y}_1 \int_{-a}^{a} d{y}_2 u_x( {y}_1)u_x({y}_2) B({y}_1\vert {y}_2)p_b({y}_2) \label{eq:DePlanar},
\end{align}
with $P_B$ the stationary probability to be observed in the bulk, $P_S$ the  stationary probability to be observed on the surfaces, and $B$ defined by 
\begin{align}
B( {y}_1\vert  {y}_2) =\frac{1}{2D_b(a+\delta)^2} \Bigg[ \frac{a^3}{3} +a^2\delta +a\delta^2 \nonumber\\
-(a+\delta)^2\vert y_1-y_2\vert+(a+\delta)\frac{(y_1^2+y_2^2)}{2}
\Bigg],  
\end{align}
with   $\delta=k_a/k_d$. Moreover, if $p_b(y,t)$ is the marginal probability to observe the particle in the bulk at position $y$ in the steady state, and $p_s^+$ and $p_s^-$ the probability to observe the particle attached to the upper (and lower) wall, respectively, one has
\begin{align}
p_b k_a =p_s^+k_d=p_s^-k_d, \ 
p_s^+=p_s^-=\frac{P_S}{2},\ 
P_B=2a p_b ,
\end{align}
from which it is easy to show that 
\begin{align}
&p_b = \frac{1}{2(a + \delta)}, &p_s^+=p_s^- = \frac{\delta}{2(a + \delta)},\\
&P_B= \frac{a}{a + \delta}, &P_S= \frac{\delta}{a + \delta}.
\end{align}
Applying \cref{eq:DePlanar} with $u_e(y)=u_x^{(0)}(y)+L u_x^{(1)}(y)$ leads to Eq.~(\ref{eq:DeAfterMap})   for the effective diffusivity in a planar channel.

\section{Diffusivity for translationally invariant channel with anisotropic, spatially varying diffusivity}
\label{sec:AnisotropicFlat}
For $d=2$, the equilibrium probability  (\ref{Def_pe}) for the lateral variable $y$ reads
\begin{align}
p_e(y)=\frac{1}{\langle h\rangle}\times
\begin{cases}
 W(y) & (\vert y\vert>a),\\
 1 & (\vert y\vert<a).
\end{cases}
\end{align}
Noting that $u_e$ vanishes for $\vert y\vert >a$, the effective diffusivity obtained with \cref{eq:DeMappingNonUniformDiffChann} reads
\begin{align}
&D_{e} = \frac{(a+\ell)D}{\langle h \rangle}+\frac{1}{D\langle h \rangle}\int_a^{h_m}\dd y \frac{\left\{ v \int_y^{h_m}\dd y_1 W(y_1)\right\}^2}{W(y)} \nonumber\\
&+\int_0^{a}\frac{\dd y }{D\langle h \rangle}\left\{ \int_y^a\dd y_1[u_e(y_1)-v]- \int_a^{h_m}\dd y_1W(y_1)v\right\}^2 \nonumber\\
&=\frac{(a+\ell)D}{\langle h \rangle}
+\frac{1}{D\langle h \rangle}\int_a^{h_m}\dd y \frac{\left\{ v \int_y^{h_m}\dd y_1 W(y_1)\right\}^2}{W(y)} \nonumber\\
& +\frac{1}{D\langle h \rangle}\int_0^{a}\dd y \left\{ \int_y^a\dd y_1[u_x(y_1)-v]- \delta v\right\}^2 
\end{align} 
Using the form $u_e=U(1-y^2/a^2)+u_s$, we finally obtain Eq.~(\ref{eq:DeAfterMap}) if one sets $\ell = L\ln 2/\pi$.


\begin{thebibliography}{10}
\expandafter\ifx\csname url\endcsname\relax
  \def\url#1{\texttt{#1}}\fi
\expandafter\ifx\csname urlprefix\endcsname\relax\def\urlprefix{URL }\fi
\providecommand{\bibinfo}[2]{#2}
\providecommand{\eprint}[2][]{\url{#2}}

\bibitem{marbach2018transport}
\bibinfo{author}{Marbach, S.}, \bibinfo{author}{Dean, D.~S.} \&
  \bibinfo{author}{Bocquet, L.}
\newblock \bibinfo{title}{Transport and dispersion across wiggling nanopores}.
\newblock \emph{\bibinfo{journal}{Nat. Phys.}} \textbf{\bibinfo{volume}{14}},
  \bibinfo{pages}{1108--1113} (\bibinfo{year}{2018}).

\bibitem{aminian2016boundaries}
\bibinfo{author}{Aminian, M.}, \bibinfo{author}{Bernardi, F.},
  \bibinfo{author}{Camassa, R.}, \bibinfo{author}{Harris, D.~M.} \&
  \bibinfo{author}{McLaughlin, R.~M.}
\newblock \bibinfo{title}{How boundaries shape chemical delivery in
  microfluidics}.
\newblock \emph{\bibinfo{journal}{Science}} \textbf{\bibinfo{volume}{354}},
  \bibinfo{pages}{1252--1256} (\bibinfo{year}{2016}).

\bibitem{kim2019tuning}
\bibinfo{author}{Kim, W.~K.}, \bibinfo{author}{Kandu{\v{c}}, M.},
  \bibinfo{author}{Roa, R.} \& \bibinfo{author}{Dzubiella, J.}
\newblock \bibinfo{title}{Tuning the permeability of dense membranes by shaping
  nanoscale potentials}.
\newblock \emph{\bibinfo{journal}{Phys. Rev. Lett.}}
  \textbf{\bibinfo{volume}{122}}, \bibinfo{pages}{108001}
  (\bibinfo{year}{2019}).

\bibitem{brenner1993macrotransport}
\bibinfo{author}{Brenner, H.} \& \bibinfo{author}{Edwards, D.~A.}
\newblock \emph{\bibinfo{title}{Macrotransport {P}rocesses, edited by
  {B}utterworth}} (\bibinfo{publisher}{Heinemann}, \bibinfo{year}{1993}).

\bibitem{meigel2022dispersive}
\bibinfo{author}{Meigel, F.~J.}, \bibinfo{author}{Darwent, T.},
  \bibinfo{author}{Bastin, L.}, \bibinfo{author}{Goehring, L.} \&
  \bibinfo{author}{Alim, K.}
\newblock \bibinfo{title}{Dispersive transport dynamics in porous media emerge
  from local correlations}.
\newblock \emph{\bibinfo{journal}{Nat. Comm.}} \textbf{\bibinfo{volume}{13}},
  \bibinfo{pages}{5885} (\bibinfo{year}{2022}).

\bibitem{le2013stretching}
\bibinfo{author}{Le~Borgne, T.}, \bibinfo{author}{Dentz, M.} \&
  \bibinfo{author}{Villermaux, E.}
\newblock \bibinfo{title}{Stretching, coalescence, and mixing in porous media}.
\newblock \emph{\bibinfo{journal}{Phys. Rev. Lett.}}
  \textbf{\bibinfo{volume}{110}}, \bibinfo{pages}{204501}
  (\bibinfo{year}{2013}).

\bibitem{dentz2011mixing}
\bibinfo{author}{Dentz, M.}, \bibinfo{author}{Le~Borgne, T.},
  \bibinfo{author}{Englert, A.} \& \bibinfo{author}{Bijeljic, B.}
\newblock \bibinfo{title}{Mixing, spreading and reaction in heterogeneous
  media: {A} brief review}.
\newblock \emph{\bibinfo{journal}{J. Contam. Hydrol.}}
  \textbf{\bibinfo{volume}{120}}, \bibinfo{pages}{1--17}
  (\bibinfo{year}{2011}).

\bibitem{de2012flow}
\bibinfo{author}{De~Barros, F.~P.}, \bibinfo{author}{Dentz, M.},
  \bibinfo{author}{Koch, J.} \& \bibinfo{author}{Nowak, W.}
\newblock \bibinfo{title}{Flow topology and scalar mixing in spatially
  heterogeneous flow fields}.
\newblock \emph{\bibinfo{journal}{Geophys. Res. Lett.}}
  \textbf{\bibinfo{volume}{39}} (\bibinfo{year}{2012}).

\bibitem{bernate2012stochastic}
\bibinfo{author}{Bernate, J.~A.} \& \bibinfo{author}{Drazer, G.}
\newblock \bibinfo{title}{Stochastic and deterministic vector chromatography of
  suspended particles in one-dimensional periodic potentials}.
\newblock \emph{\bibinfo{journal}{Phys. Rev. Lett.}}
  \textbf{\bibinfo{volume}{108}}, \bibinfo{pages}{214501}
  (\bibinfo{year}{2012}).

\bibitem{aminian2015squaring}
\bibinfo{author}{Aminian, M.}, \bibinfo{author}{Bernardi, F.},
  \bibinfo{author}{Camassa, R.} \& \bibinfo{author}{McLaughlin, R.~M.}
\newblock \bibinfo{title}{Squaring the circle: {G}eometric skewness and
  symmetry breaking for passive scalar transport in ducts and pipes}.
\newblock \emph{\bibinfo{journal}{Phys. Rev. Lett.}}
  \textbf{\bibinfo{volume}{115}}, \bibinfo{pages}{154503}
  (\bibinfo{year}{2015}).

\bibitem{dean2007effective}
\bibinfo{author}{Dean, D.~S.}, \bibinfo{author}{Drummond, I.} \&
  \bibinfo{author}{Horgan, R.}
\newblock \bibinfo{title}{Effective transport properties for diffusion in
  random media}.
\newblock \emph{\bibinfo{journal}{J. Stat. Mech: Theor. Exp.}}
  \textbf{\bibinfo{volume}{2007}}, \bibinfo{pages}{P07013}
  (\bibinfo{year}{2007}).

\bibitem{malgaretti2013entropic}
\bibinfo{author}{Malgaretti, P.}, \bibinfo{author}{Pagonabarraga, I.} \&
  \bibinfo{author}{Rubi, J.~M.}
\newblock \bibinfo{title}{Entropic transport in confined media: a challenge for
  computational studies in biological and soft-matter systems}.
\newblock \emph{\bibinfo{journal}{Front. Phys.}} \textbf{\bibinfo{volume}{1}},
  \bibinfo{pages}{21} (\bibinfo{year}{2013}).

\bibitem{burada2009diffusion}
\bibinfo{author}{Burada, P.~S.}, \bibinfo{author}{H{\"a}nggi, P.},
  \bibinfo{author}{Marchesoni, F.}, \bibinfo{author}{Schmid, G.} \&
  \bibinfo{author}{Talkner, P.}
\newblock \bibinfo{title}{Diffusion in confined geometries}.
\newblock \emph{\bibinfo{journal}{ChemPhysChem}} \textbf{\bibinfo{volume}{10}},
  \bibinfo{pages}{45--54} (\bibinfo{year}{2009}).

\bibitem{jacobs1967diffusion}
\bibinfo{author}{Jacobs, M.}
\newblock \bibinfo{title}{Diffusion {P}rocesses; springer}.
\newblock \emph{\bibinfo{journal}{Berlin Heidelberg New York}}
  (\bibinfo{year}{1967}).

\bibitem{reguera2006entropic}
\bibinfo{author}{Reguera, D.} \emph{et~al.}
\newblock \bibinfo{title}{Entropic transport: {K}inetics, scaling, and control
  mechanisms}.
\newblock \emph{\bibinfo{journal}{Phys. Rev. Lett.}}
  \textbf{\bibinfo{volume}{96}}, \bibinfo{pages}{130603}
  (\bibinfo{year}{2006}).

\bibitem{rubi2019entropic}
\bibinfo{author}{Rubi, J.~M.}
\newblock \bibinfo{title}{Entropic diffusion in confined soft-matter and
  biological systems}.
\newblock \emph{\bibinfo{journal}{Europhys. Lett.}}
  \textbf{\bibinfo{volume}{127}}, \bibinfo{pages}{10001}
  (\bibinfo{year}{2019}).

\bibitem{taylor1953dispersion}
\bibinfo{author}{Taylor, G.~I.}
\newblock \bibinfo{title}{Dispersion of soluble matter in solvent flowing
  slowly through a tube}.
\newblock \emph{\bibinfo{journal}{Proc. R. Soc. Lond., A}}
  \textbf{\bibinfo{volume}{219}}, \bibinfo{pages}{186--203}
  (\bibinfo{year}{1953}).

\bibitem{martens2013hydrodynamically}
\bibinfo{author}{Martens, S.}, \bibinfo{author}{Straube, A.},
  \bibinfo{author}{Schmid, G.}, \bibinfo{author}{Schimansky-Geier, L.} \&
  \bibinfo{author}{H{\"a}nggi, P.}
\newblock \bibinfo{title}{Hydrodynamically enforced entropic trapping of
  {B}rownian particles}.
\newblock \emph{\bibinfo{journal}{Phys. Rev. Lett.}}
  \textbf{\bibinfo{volume}{110}}, \bibinfo{pages}{010601}
  (\bibinfo{year}{2013}).

\bibitem{kalinay2020taylor}
\bibinfo{author}{Kalinay, P.}
\newblock \bibinfo{title}{Taylor dispersion in {P}oiseuille flow in
  three-dimensional tubes of varying diameter}.
\newblock \emph{\bibinfo{journal}{Phys. Rev. E}}
  \textbf{\bibinfo{volume}{102}}, \bibinfo{pages}{042606}
  (\bibinfo{year}{2020}).

\bibitem{bolster2009solute}
\bibinfo{author}{Bolster, D.}, \bibinfo{author}{Dentz, M.} \&
  \bibinfo{author}{Le~Borgne, T.}
\newblock \bibinfo{title}{Solute dispersion in channels with periodically
  varying apertures}.
\newblock \emph{\bibinfo{journal}{Phys. Fluids}} \textbf{\bibinfo{volume}{21}}
  (\bibinfo{year}{2009}).

\bibitem{yang2017hydrodynamic}
\bibinfo{author}{Yang, X.} \emph{et~al.}
\newblock \bibinfo{title}{Hydrodynamic and entropic effects on colloidal
  diffusion in corrugated channels}.
\newblock \emph{\bibinfo{journal}{Proc. Natl. Acad. Sci. U.S.A.}}
  \textbf{\bibinfo{volume}{114}}, \bibinfo{pages}{9564--9569}
  (\bibinfo{year}{2017}).

\bibitem{smith1983longitudinal}
\bibinfo{author}{Smith, R.}
\newblock \bibinfo{title}{Longitudinal dispersion coefficients for varying
  channels}.
\newblock \emph{\bibinfo{journal}{J. Fluid. Mech.}}
  \textbf{\bibinfo{volume}{130}}, \bibinfo{pages}{299--314}
  (\bibinfo{year}{1983}).

\bibitem{hoagland1985taylor}
\bibinfo{author}{Hoagland, D.} \& \bibinfo{author}{Prud'Homme, R.}
\newblock \bibinfo{title}{Taylor-{A}ris dispersion arising from flow in a
  sinusoidal tube}.
\newblock \emph{\bibinfo{journal}{AIChE journal}}
  \textbf{\bibinfo{volume}{31}}, \bibinfo{pages}{236--244}
  (\bibinfo{year}{1985}).

\bibitem{marbach2019active}
\bibinfo{author}{Marbach, S.} \& \bibinfo{author}{Alim, K.}
\newblock \bibinfo{title}{Active control of dispersion within a channel with
  flow and pulsating walls}.
\newblock \emph{\bibinfo{journal}{Phys. Rev. Fluids}}
  \textbf{\bibinfo{volume}{4}}, \bibinfo{pages}{114202} (\bibinfo{year}{2019}).

\bibitem{dagdug2012projection}
\bibinfo{author}{Dagdug, L.} \& \bibinfo{author}{Pineda, I.}
\newblock \bibinfo{title}{Projection of two-dimensional diffusion in a curved
  midline and narrow varying width channel onto the longitudinal dimension}.
\newblock \emph{\bibinfo{journal}{J. Chem. Phys.}}
  \textbf{\bibinfo{volume}{137}}, \bibinfo{pages}{024107}
  (\bibinfo{year}{2012}).

\bibitem{roggeveen2023transport}
\bibinfo{author}{Roggeveen, J.~V.}, \bibinfo{author}{Stone, H.~A.} \&
  \bibinfo{author}{Kurzthaler, C.}
\newblock \bibinfo{title}{Transport of a passive scalar in wide channels with
  surface topography: {A}n asymptotic theory}.
\newblock \emph{\bibinfo{journal}{J. Phys.: Condens. Matter}}
  \textbf{\bibinfo{volume}{35}}, \bibinfo{pages}{274003}
  (\bibinfo{year}{2023}).

\bibitem{haugerud2022solute}
\bibinfo{author}{Haugerud, I.~S.}, \bibinfo{author}{Linga, G.} \&
  \bibinfo{author}{Flekk{\o}y, E.~G.}
\newblock \bibinfo{title}{Solute dispersion in channels with periodic square
  boundary roughness}.
\newblock \emph{\bibinfo{journal}{J. Fluid. Mech.}}
  \textbf{\bibinfo{volume}{944}}, \bibinfo{pages}{A53} (\bibinfo{year}{2022}).

\bibitem{amaral1997dispersion_part1}
\bibinfo{author}{Amaral~Souto, H.~P.} \& \bibinfo{author}{Moyne, C.}
\newblock \bibinfo{title}{Dispersion in two-dimensional periodic porous media.
  {P}art ii. {D}ispersion tensor}.
\newblock \emph{\bibinfo{journal}{Phys. Fluids}} \textbf{\bibinfo{volume}{9}},
  \bibinfo{pages}{2253--2263} (\bibinfo{year}{1997}).

\bibitem{amaral1997dispersion_part2}
\bibinfo{author}{Amaral~Souto, H.~P.} \& \bibinfo{author}{Moyne, C.}
\newblock \bibinfo{title}{Dispersion in two-dimensional periodic porous media.
  {P}art i. {H}ydrodynamics}.
\newblock \emph{\bibinfo{journal}{Phys. Fluids}} \textbf{\bibinfo{volume}{9}},
  \bibinfo{pages}{2243--2252} (\bibinfo{year}{1997}).

\bibitem{quintard1994convection}
\bibinfo{author}{Quintard, M.} \& \bibinfo{author}{Whitaker, S.}
\newblock \bibinfo{title}{Convection, dispersion, and interfacial transport of
  contaminants: {H}omogeneous porous media}.
\newblock \emph{\bibinfo{journal}{Adv. Water Resour.}}
  \textbf{\bibinfo{volume}{17}}, \bibinfo{pages}{221--239}
  (\bibinfo{year}{1994}).

\bibitem{quintard1993transport}
\bibinfo{author}{Quintard, M.} \& \bibinfo{author}{Whitaker, S.}
\newblock \bibinfo{title}{Transport in ordered and disordered porous media:
  volume-averaged equations, closure problems, and comparison with experiment}.
\newblock \emph{\bibinfo{journal}{Chem. Eng. Sci.}}
  \textbf{\bibinfo{volume}{48}}, \bibinfo{pages}{2537--2564}
  (\bibinfo{year}{1993}).

\bibitem{richardson1973no}
\bibinfo{author}{Richardson, S.}
\newblock \bibinfo{title}{On the no-slip boundary condition}.
\newblock \emph{\bibinfo{journal}{J. Fluid Mech.}}
  \textbf{\bibinfo{volume}{59}}, \bibinfo{pages}{707--719}
  (\bibinfo{year}{1973}).

\bibitem{jeong2001slip}
\bibinfo{author}{Jeong, J.-T.}
\newblock \bibinfo{title}{Slip boundary condition on an idealized porous wall}.
\newblock \emph{\bibinfo{journal}{Phys. fluids}} \textbf{\bibinfo{volume}{13}},
  \bibinfo{pages}{1884--1890} (\bibinfo{year}{2001}).

\bibitem{jansons1988determination}
\bibinfo{author}{Jansons, K.~M.}
\newblock \bibinfo{title}{Determination of the macroscopic (partial) slip
  boundary condition for a viscous flow over a randomly rough surface with a
  perfect slip microscopic boundary condition}.
\newblock \emph{\bibinfo{journal}{Phys. Fluids}} \textbf{\bibinfo{volume}{31}},
  \bibinfo{pages}{15--17} (\bibinfo{year}{1988}).

\bibitem{lauga2005microfluidics}
\bibinfo{author}{Lauga, E.}, \bibinfo{author}{Brenner, M.~P.} \&
  \bibinfo{author}{Stone, H.~A.}
\newblock \bibinfo{title}{Microfluidics: the no-slip boundary condition}.
\newblock \emph{\bibinfo{journal}{In ``Handbook of Experimental Fluid
  Dynamics'', C. Tropea, A. Yarin, J.F. Foss (Eds)}}  (\bibinfo{year}{2007}).

\bibitem{bocquet2010nanofluidics}
\bibinfo{author}{Bocquet, L.} \& \bibinfo{author}{Charlaix, E.}
\newblock \bibinfo{title}{Nanofluidics, from bulk to interfaces}.
\newblock \emph{\bibinfo{journal}{Chem. Soc. Rev.}}
  \textbf{\bibinfo{volume}{39}}, \bibinfo{pages}{1073--1095}
  (\bibinfo{year}{2010}).

\bibitem{squires2005microfluidics}
\bibinfo{author}{Squires, T.~M.} \& \bibinfo{author}{Quake, S.~R.}
\newblock \bibinfo{title}{Microfluidics: {F}luid physics at the nanoliter
  scale}.
\newblock \emph{\bibinfo{journal}{Rev. Mod. Phys.}}
  \textbf{\bibinfo{volume}{77}}, \bibinfo{pages}{977--1026}
  (\bibinfo{year}{2005}).

\bibitem{benAvraham2000}
\bibinfo{author}{ben Avraham, D.} \& \bibinfo{author}{Havlin, S.}
\newblock \emph{\bibinfo{title}{Diffusion and reactions in Fractals and
  Disordered systems}} (\bibinfo{publisher}{Cambridge University Press,
  Cambridge, UK}, \bibinfo{year}{2000}).

\bibitem{dagdug2014aris}
\bibinfo{author}{Dagdug, L.}, \bibinfo{author}{Berezhkovskii, A.~M.} \&
  \bibinfo{author}{Skvortsov, A.~T.}
\newblock \bibinfo{title}{Aris-{T}aylor dispersion in tubes with dead ends}.
\newblock \emph{\bibinfo{journal}{J. Chem. Phys.}}
  \textbf{\bibinfo{volume}{141}} (\bibinfo{year}{2014}).

\bibitem{berezhkovskii2014normal}
\bibinfo{author}{Berezhkovskii, A.~M.}, \bibinfo{author}{Dagdug, L.} \&
  \bibinfo{author}{Bezrukov, S.~M.}
\newblock \bibinfo{title}{From normal to anomalous diffusion in comb-like
  structures in three dimensions}.
\newblock \emph{\bibinfo{journal}{J. Chem. Phys.}}
  \textbf{\bibinfo{volume}{141}}, \bibinfo{pages}{054907}
  (\bibinfo{year}{2014}).

\bibitem{bettarini2024effective}
\bibinfo{author}{Bettarini, G.} \& \bibinfo{author}{Piazza, F.}
\newblock \bibinfo{title}{Effective diffusion along the backbone of combs with
  finite-span 1d and 2d fingers}.
\newblock \emph{\bibinfo{journal}{J. Chem. Phys.}}
  \textbf{\bibinfo{volume}{161}} (\bibinfo{year}{2024}).

\bibitem{rubinstein1986dispersion}
\bibinfo{author}{Rubinstein, J.} \& \bibinfo{author}{Mauri, R.}
\newblock \bibinfo{title}{Dispersion and convection in periodic porous media}.
\newblock \emph{\bibinfo{journal}{SIAM J. Appl. Math.}}
  \textbf{\bibinfo{volume}{46}}, \bibinfo{pages}{1018--1023}
  (\bibinfo{year}{1986}).

\bibitem{alshare2010modeling}
\bibinfo{author}{Alshare, A.}, \bibinfo{author}{Strykowski, P.~J.} \&
  \bibinfo{author}{Simon, T.~W.}
\newblock \bibinfo{title}{Modeling of unsteady and steady fluid flow, heat
  transfer and dispersion in porous media using unit cell scale}.
\newblock \emph{\bibinfo{journal}{Int. J. Heat Mass Transfer}}
  \textbf{\bibinfo{volume}{53}}, \bibinfo{pages}{2294--2310}
  (\bibinfo{year}{2010}).

\bibitem{brenner1980dispersion}
\bibinfo{author}{Brenner, H.}
\newblock \bibinfo{title}{Dispersion resulting from flow through spatially
  periodic porous media}.
\newblock \emph{\bibinfo{journal}{Philosophical Transactions of the Royal
  Society of London. Series A, Mathematical and Physical Sciences}}
  \textbf{\bibinfo{volume}{297}}, \bibinfo{pages}{81--133}
  (\bibinfo{year}{1980}).

\bibitem{guerin2015kubo}
\bibinfo{author}{Gu{\'e}rin, T.} \& \bibinfo{author}{Dean, D.~S.}
\newblock \bibinfo{title}{Kubo formulas for dispersion in heterogeneous
  periodic nonequilibrium systems}.
\newblock \emph{\bibinfo{journal}{Phys. Rev. E}} \textbf{\bibinfo{volume}{92}},
  \bibinfo{pages}{062103} (\bibinfo{year}{2015}).

\bibitem{guerin2015force}
\bibinfo{author}{Gu{\'e}rin, T.} \& \bibinfo{author}{Dean, D.~S.}
\newblock \bibinfo{title}{Force-induced dispersion in heterogeneous media}.
\newblock \emph{\bibinfo{journal}{Phys. Rev. Lett.}}
  \textbf{\bibinfo{volume}{115}}, \bibinfo{pages}{020601}
  (\bibinfo{year}{2015}).

\bibitem{supmat}
\bibinfo{title}{See {S}upplemental {M}aterial at [url will be inserted by
  publisher] for calculation details} .

\bibitem{ward1993strong}
\bibinfo{author}{Ward, M.~J.} \& \bibinfo{author}{Keller, J.~B.}
\newblock \bibinfo{title}{Strong localized perturbations of eigenvalue
  problems}.
\newblock \emph{\bibinfo{journal}{SIAM J. Appl. Math.}}
  \textbf{\bibinfo{volume}{53}}, \bibinfo{pages}{770--798}
  (\bibinfo{year}{1993}).

\bibitem{luchini1991resistance}
\bibinfo{author}{Luchini, P.}, \bibinfo{author}{Manzo, F.} \&
  \bibinfo{author}{Pozzi, A.}
\newblock \bibinfo{title}{Resistance of a grooved surface to parallel flow and
  cross-flow}.
\newblock \emph{\bibinfo{journal}{J. Fluid. Mech.}}
  \textbf{\bibinfo{volume}{228}}, \bibinfo{pages}{87--109}
  (\bibinfo{year}{1991}).

\bibitem{mangeat2017geometry}
\bibinfo{author}{Mangeat, M.}, \bibinfo{author}{Gu{\'e}rin, T.} \&
  \bibinfo{author}{Dean, D.~S.}
\newblock \bibinfo{title}{Geometry controlled dispersion in periodic corrugated
  channels}.
\newblock \emph{\bibinfo{journal}{Europhys. Lett.}}
  \textbf{\bibinfo{volume}{118}}, \bibinfo{pages}{40004}
  (\bibinfo{year}{2017}).

\bibitem{levesque2012taylor}
\bibinfo{author}{Levesque, M.}, \bibinfo{author}{B{\'e}nichou, O.},
  \bibinfo{author}{Voituriez, R.} \& \bibinfo{author}{Rotenberg, B.}
\newblock \bibinfo{title}{Taylor dispersion with adsorption and desorption}.
\newblock \emph{\bibinfo{journal}{Phys. Rev. E}} \textbf{\bibinfo{volume}{86}},
  \bibinfo{pages}{036316} (\bibinfo{year}{2012}).

\bibitem{venditti2022exact}
\bibinfo{author}{Venditti, C.}, \bibinfo{author}{Giona, M.} \&
  \bibinfo{author}{Adrover, A.}
\newblock \bibinfo{title}{Exact moment analysis of transient/asymptotic
  dispersion properties in periodic media with adsorbing/desorbing walls}.
\newblock \emph{\bibinfo{journal}{Phys. Fluids.}} \textbf{\bibinfo{volume}{34}}
  (\bibinfo{year}{2022}).

\bibitem{jiang2022analytical}
\bibinfo{author}{Jiang, W.}, \bibinfo{author}{Zeng, L.}, \bibinfo{author}{Fu,
  X.} \& \bibinfo{author}{Wu, Z.}
\newblock \bibinfo{title}{Analytical solutions for reactive shear dispersion
  with boundary adsorption and desorption}.
\newblock \emph{\bibinfo{journal}{J. Fluid. Mech.}}
  \textbf{\bibinfo{volume}{947}}, \bibinfo{pages}{A37} (\bibinfo{year}{2022}).

\bibitem{hlushkou2014effect}
\bibinfo{author}{Hlushkou, D.}, \bibinfo{author}{Gritti, F.},
  \bibinfo{author}{Guiochon, G.}, \bibinfo{author}{Seidel-Morgenstern, A.} \&
  \bibinfo{author}{Tallarek, U.}
\newblock \bibinfo{title}{Effect of adsorption on solute dispersion: a
  microscopic stochastic approach}.
\newblock \emph{\bibinfo{journal}{Analytical chemistry}}
  \textbf{\bibinfo{volume}{86}}, \bibinfo{pages}{4463--4470}
  (\bibinfo{year}{2014}).

\bibitem{berezhkovskii2013aris}
\bibinfo{author}{Berezhkovskii, A.~M.} \& \bibinfo{author}{Skvortsov, A.~T.}
\newblock \bibinfo{title}{Aris-{T}aylor dispersion with drift and diffusion of
  particles on the tube wall}.
\newblock \emph{\bibinfo{journal}{J. Chem. Phys.}}
  \textbf{\bibinfo{volume}{139}} (\bibinfo{year}{2013}).

\bibitem{szabo1980first}
\bibinfo{author}{Szabo, A.}, \bibinfo{author}{Schulten, K.} \&
  \bibinfo{author}{Schulten, Z.}
\newblock \bibinfo{title}{First passage time approach to diffusion controlled
  reactions}.
\newblock \emph{\bibinfo{journal}{J. Chem. Phys.}}
  \textbf{\bibinfo{volume}{72}}, \bibinfo{pages}{4350--4357}
  (\bibinfo{year}{1980}).

\bibitem{alexandre2021generalized}
\bibinfo{author}{Alexandre, A.}, \bibinfo{author}{Gu{\'e}rin, T.} \&
  \bibinfo{author}{Dean, D.~S.}
\newblock \bibinfo{title}{Generalized {T}aylor dispersion for translationally
  invariant microfluidic systems}.
\newblock \emph{\bibinfo{journal}{Phys. Fluids.}} \textbf{\bibinfo{volume}{33}}, \bibinfo{pages}{082004}
  (\bibinfo{year}{2021})

\bibitem{bechert1989viscous}
\bibinfo{author}{Bechert, D.W.},   \&
  \bibinfo{author}{Bartenwerfer, M.}
\newblock \bibinfo{title}{The viscous flow on surfaces with longitudinal ribs}.
\newblock \emph{\bibinfo{journal}{J. Fluid. Mech.}} \textbf{\bibinfo{volume}{206}}, \bibinfo{pages}{105--129}
  (\bibinfo{year}{1989}).
  
  \bibitem{Alexandre2025DATA}
 \bibinfo{author}{Alexandre, A.}, \bibinfo{author}{Gu{\'e}rin, T.} \&
  \bibinfo{author}{Dean, D.~S.}
Data for ``Effective description of Taylor dispersion in strongly corrugated channels'' [Data set]. Zenodo. https://doi.org/10.5281/zenodo.15425811 (2025). 



\end{thebibliography}

\section{Calculation details in 3D for axisymmetric channels}
\label{sec:3DTheory}
Here we explain how to generalize the calculation of the diffusivity to the case of a three-dimensional axisymmetric channel. The calculations here are essentially the same as those of Section \cref{sec:CalculDe} but are presented in detail for the sake of clarity. Here we keep the notation $x$ to represent the coordinate along the channel axis, $x=X/L$, $y>0$ denotes the distance to the central axis, and $h(X)$ is the local channel radius. Let us first consider the pressure-induced flow. As in 2D, the structure of the flow is  given by
\begin{align}
& \ve[u] \underset{L\to0}{\simeq}
\begin{cases}
\ve[u]_0(X,y)+L \ve[u]_1(X,y)+...& [y<a],\\
L\,\ve[u]^*(X,Y)+...& [y-a=\mathcal{O}(L)],\\
0 & [y>a], 
\end{cases}\label{eq:ExpFlow2}
\end{align}
At leading order, the flow $\ve[u]_0$ is a pressure induced flow in a uniform channel of radius $a$, since it depends only on $y$ the Stokes equations take the form
\begin{align}
\eta \left(\frac{1}{y}\partial_y (y \partial_y u_x)+\partial_x^2 u_x \right) -\partial_x \Pi=0.
\end{align}
This equation, with the boundary condition $\ve[u]_0(y=a)=0$ is readily solved:
\begin{align}
&\ve[u]_0=U(1-y^2/a^2) \,{\ve[e]}_x, &U=-\frac{a^2 (\nabla \Pi)_\infty}{2\eta}.
\end{align}
Next, near the entrance of the lateral regions, the field $\ve[u]^*(X,Y)$ satisfies exactly the same equations as in the two-dimensional case, so that the effective slip velocity is still given by $u_s=UL\beta/a$. This leads to a flow at next-to-leading order which is uniform
\begin{align}
\ve[u]_1= (U\beta/a  )\,{\ve[e]}_x. 
\end{align}
As a consequence the average flow in the $x$ direction reads
\begin{align}
v =  \frac{\int_0^a \dd y y [U(1-y^2/a^2)+UL\beta/a] }{\int_{-1/2}^{1/2} \dd X \int_0^{h(X)}\dd y y } =  v_0+Lv_1+\mathcal{O}(L^2),
\end{align}
with
\begin{align}
&v_0=\frac{a^2 U}{2\langle h^2\rangle},&v_1=\frac{U a\beta }{\langle h^2\rangle}.\label{eq:v0v13D}
\end{align}

Next, we focus on the calculation of $f$, which satisfies the equations
\begin{align}
-\frac{u_x}{L} \partial_X f  + D\left(\frac{1}{L^2} \partial_X^2f+\frac{1}{y} \partial_y  (y\partial_yf)\right) = v-u_x , \label{eq:EqfRescaled3D}
\end{align} 
in both peripheral and central regions. In the peripheral region, the boundary condition is
\begin{align}
& \frac{1}{L}(\partial_X h) \left(\frac{\partial_Xf^p}{L}-1\right)-\partial_yf^p =0 & [y=h(X)] \label{eq:BCRescaled3D},
\end{align}
while for the central region one has the condition that $f$ is periodic of period $1$ and the condition $\partial_yf^c=0$ at $y=0$. 

Inserting the series expansion in powers of $L$, $f^w=\sum_{n\ge0} L^n f_n^w(X,y)$ ($w \in \left\{c,p\right\}$) in both central and peripheral regions into \cref{eq:EqfRescaled3D,eq:BCRescaled3D}, we find  that \cref{eq:Eqf0,eq:Eqf1,eq:BC0,eq:BC1} found for the two-dimensional case are unchanged, so that the general form of solutions for $f$ is unchanged at order $1$ and $L$:
\begin{align} 
&f_0^{c}(X,y)=f_0^{c}(y) , & f_0^{p}(X,y)=f_0^{p}(y), \\
&f_1^{c}(X,y)=f_1^{c}(y) , & f_1^{p}(X,y)=X+b_1^p(y),
\end{align}
where $f_0^c,f_0^p,f_1^c,b_1^p$ are functions of $y$ only. 

To determine these functions, we need to consider the next orders in the power expansion of  \cref{eq:EqfRescaled3D,eq:BCRescaled3D}. First, for the peripheral region (where $u_x=0$), we obtain
\begin{align}
&D\left[\partial_X^2 f_2^p+\frac{1}{y}\partial_y (y \partial_y f_0^p)\right]  =v_0, \\
&D\left[\partial_X^2 f_3^p+\frac{1}{y}\partial_y(y\partial_y f_1^p)\right] = v_1, 
\end{align}
Integrating over the variable $X$, we obtain
\begin{align}
&D\partial_X f_2^p  =\left[v_0-D\frac{1}{y}\partial_y (y \partial_y f_0^p(y))\right]X+A_2^p(y), \label{eq:Eqdxf23D}\\
&D\partial_X f_3^p  =\left[v_1-D\frac{1}{y}\partial_y (y \partial_y b_1^p(y))\right]X+A_3^p(y),\label{eq:Eqdxf33D} 
\end{align}
where $A_2^p$ and $A_3^p$ are functions of $y$ only.  The boundary conditions are
\begin{align}
&(h'(X)\partial_X f_2^p-\partial_yf_0^p)_{y=h(X)}=0 ,\\
&(h'(X)\partial_X f_3^p-\partial_yf_1^p)_{y=h(X)}=0,
\end{align}
so that, using \cref{eq:Eqdxf23D,eq:Eqdxf33D}, we obtain
\begin{align}
h'(X)\Bigg\{&\left[v_0-D  (f_0^p)''(h(X))-D\frac{(f_0^p)'(h(X))}{h(X)}\right]X\nonumber\\
&+A_2^p (h(X)) \Bigg\}= D (f_0^p)'(h(X)), \\
h'(X)\Bigg\{&\left[v_1-D  (b_1^p)''(h(X))-D\frac{(b_1^p)'(h(X))}{h(X)})\right]X\nonumber\\
&+A_3^p (h(X))\Bigg\} = D (b_1^p)'(h(X)). 
\end{align}
Multiplying these equations by $h(X)$ and integrating once leads to
\begin{align}
D X h(X)(f_0^p)'(h(X)) =
v_0 \int_{-1/2}^X \dd w\,h'(w)h(w)  w \nonumber\\+ G_2^p(h(X))  \label{eq:054923D}, \\
 D Xh(X)(b_1^p)'(h(X)) =v_1 \int_{-1/2}^X \dd w \,h'(w)h(w)  w \nonumber\\+ G_3^p(h(X)),
\end{align}
where $G_2^p$ and $G_3^p$ are primitive functions of $yA_2^p$ and $yA_3^p$, respectively: $\partial_y[G_i^p(y)]=yA_i^p(y)$ ($i\in \left\{2,3 \right\}$).

Now, for a given value of $y$, we can find two values of $X$, say $X_1(y)$ and $X_2(y)$ so that $y=h(X_1(y))=h(X_2(y))$, with $X_1(y)<X_2(y)$. 
If we write the above equations for $X_1(y)$ and $X_2(y)$ and take the difference between the two, we obtain
\begin{align}
& D[X_2(y)-X_1(y)]\,y \,(f_0^p)'(y) =v_0  \int_{X_1(y)}^{X_2(y)}\dd w \,w\,h(w) h'(w), \label{eq:eq_f0p3D} \\
& D [X_2(y)-X_1(y)]\,y \,(b_1^p)'(y) =v_1 \int_{X_1(y)}^{X_2(y)}\dd w \,w\,h(w)  h'(w).\label{eq:eq_b1p3D}
\end{align}
Using the change of variable $y_1=h(w)$ in the above integrals (after having separated the integration interval between the intervals $[X_1(y);X_1(h_m)]$ and $[X_2(h_m);X_2(y)]$, we remark that
\begin{align}
\int_{X_1(y)}^{X_2(y)}\dd w \, w \, h(w) h'(w) =- \int_y^{h_m}\dd y_1 y_1W(y_1),
\end{align}
where $W(y)=X_2(y)-X_1(y)$ as before. We thus obtain 
 \begin{align}
& D (f_0^p)'(y) =- \frac{v_0}{ y \, W(y)}\int_y^{h_m}\dd y_1 y_1 W(y_1)  \label{eq:Res_f0p3D}, \\
& D (b_1^p)'(y) =- \frac{v_1}{y \, W(y)}\int_y^{h_m}\dd y_1y_1 W(y_1) \label{eq:Res_b1p3D}.
\end{align}

For the central region, we have 
\begin{align}
&D\left[\partial_X^2 f_2^c+\frac{1}{y}\partial_y (y\partial_yf_0^c(y))\right]  =v_0-u_x^{(0)}(y),\\
&D\left[\partial_X^2 f_3^c+\frac{1}{y}\partial_y (y\partial_yf_1^c(y))\right]  =v_1-u_x^{(1)}(y).
\end{align}
The fact that $f_2^c$ and $f_3^c$ are periodic functions of $X$ then imposes
\begin{align}
&D\frac{1}{y}\partial_y (y\partial_yf_0^c(y))=v_0-u_x^{(0)}(y)=v_0-U(1-y^2/a^2),\\
&D\frac{1}{y}\partial_y (y\partial_yf_1^c(y))=v_1-u_x^{(1)}(y)=v_1-U\beta/a.
\end{align}
Therefore, noting that $\partial_y f =0$ at $y=0$ (by symmetry, at all orders), we have
\begin{align}
&D\partial_y f_0^c (y)=v_0 \frac{y}{2} - U\left(\frac{y}{2}-\frac{y^3}{4a^2}\right)\label{eq:Res_f0c3D},\\
&D\partial_y f_1^c (y)=\left(v_1-\frac{U  \beta}{a}\right)\frac{y}{2} \label{eq:Res_f1c3D}.
\end{align}
We recall that $v_0$ and $v_1$ are given by \cref{eq:v0v13D} and that the volume of the lateral region $V_L$ can be written  as
\begin{align}
V_L=\pi L(\langle h^2\rangle -a^2)=\frac{2\pi L}{W(a)}\int_a^{h_m}dy_1 W(y_1) y_1 \nonumber\\
= 2\pi L\int_a^{h_m}dy_1 W(y_1) y_1 ,
\end{align}
with $W(a) =1$. Using these values, we see  by comparing \cref{eq:Res_f0c3D,eq:Res_f0p3D} that $\partial_y [f_0^c-f_0^p]_{y=a}=0$, so there is no discontinuity of derivative of $f_0$ at $y=a$ at this order. The same property holds for $f_1$. We can thus impose that $f_0$ is regular (continuous; with continuous derivative) at $y=a$, leading to 
\begin{align}
&Df_0(y)=\nonumber\\
&\begin{cases}
(v_0-U) (y^2-a^2)/4 + U \frac{(y^4-a^4)}{16 a^2}+C_0,& (y<a),\\
-v_0 \int_a^y \frac{dy'}{W(y')y'} \int_{y'}^{h_m} dy'' W(y'')y'' +  C_0, & (y>a),
\end{cases}\label{eq:Resultf03D}
\end{align}
where the integration constant $C_0$ is fixed by the normalization condition $\int_\Omega \dd\ve[r] f =0$, which gives
\begin{align}
\int_0^a dy \,y f_0^c(y)+\int_a^{h_m}dy\,y W(y)   f_0^p(y)=0.
\end{align}
Using the previous expressions for $f_0^c,f_0^p$, the above equation  leads to
\begin{align}
C_0  =&-\frac{2}{ \langle h^2\rangle } \Bigg\{U\frac{a^4}{24}-\frac{a^4 v_0}{16} \nonumber\\
&- v_0 \int_a^{h_m}dy\frac{\left[\int_y^{h_m}dy'' y'' W(y'')\right]^2}{W(y)y}  \Bigg\},\label{eq:ValueC03D}
\end{align}
where we have used the fact that
\begin{align}
\int_0^a dy \,y &C_0+\int_a^{h_m} dy \,y W(y) C_0 \nonumber\\
&=\int_{-1/2}^{1/2}\dd X \int_0^{h(X)}dy \,y C_0\nonumber\\
&= \int_{-1/2}^{1/2}\dd X \, \frac{h^2(X)}{2}C_0=\frac{C_0\langle h^2(X)\rangle}{2}.
\end{align}

At the boundary layer $y\simeq a$ we can perform the exact same analysis as in the 2D case, we find that $f^*$ has exactly the same form as in 2D, with the consequence that solutions can be matched only if 
\begin{align}
b_1^p(a)=f_1^c(a).
\end{align}
Applying this matching condition, we   obtain 
\begin{align}
&D f_1^c (y)=\left(v_1-\frac{U\beta}{a}\right)\frac{y^2-a^2}{4}+C_1, \label{eq:Resultf1c3D}\\
&D b_1^p(y)=-v_1 \int_a^y \frac{dy'}{W(y')y'}\int_{y'}^{h_m}dy'' W(y'')y''  + C_1,
\end{align}
and the constant $C_1$ is found by requiring that $\int_V f \dd\ve[r]=0$ at order $L$, so that
\begin{align}
\int_0^a dy &\,y D f_1^c(y)\nonumber\\
&+\int_{-1/2}^{1/2} \dd X \int_a^{h(X)} dy \,y [Db_1^p(y)+DX]=0.
\end{align}
This leads to
\begin{align}
C_1 =-\frac{2}{\langle h^2\rangle } \Bigg\{-\left(v_1-\frac{U\beta}{a}\right)  \frac{a^4}{16}  +\frac{D\langle X h^2(X) \rangle}{2} \nonumber\\
-v_1 \int_a^{h_m} \frac{dy'}{W(y') y'}\left[\int_{y'}^{h_m}dy'' W(y'')y'' \right]^2  \Bigg\}.
\end{align}
 At this stage, we have fully determined $f_0$ and $f_1$ in all the regions of the channel. To compute the diffusivity at next-to-leading order in $L$, we also need to compute $\partial_X f_2$ at the channel boundary, this can be obtained from \cref{eq:Eqdxf23D} (up to a constant) as
\begin{align}
&D\partial_X f_2^p =\left[v_0-D\frac{1}{y}\partial_y (y \partial_y f_0^p(y))\right]X+\frac{1}{y}\partial_y G_2^p(y). \label{eq:f2p3D}
\end{align}

In 3D, the effective diffusivity $D_e$ is estimated as follows:
\begin{align}
D_e=&D+ \frac{1}{ \Omega}\int_\Omega \dd\ve[r] \, [u_x f-D\partial_x f]\nonumber\\
=&D+ \frac{2}{\langle h^2\rangle}\int_{-1/2}^{1/2}\dd X \int_0^{h(X)} dy \,y \left[u_x f-\frac{D}{L}\partial_X f\right].
\end{align}
As in 2D, we write the general expansion
\begin{align}
D_e=D_e^{(0)}+L [D_A+D_B+D_C]+\mathcal{O}(L^2),
\end{align}
where the leading-order term is given by
\begin{align}
D_e^{(0)}=D+ \frac{2}{\langle h^2 \rangle}&\int_{-1/2}^{1/2}\dd X \Bigg[ \int_0^{a} dy\, y u_x^{(0)} f_0^c  \nonumber\\
& -D \int_{a}^{h(X)}dy \,y \partial_X f_1^p   \Bigg],
\end{align}
where the above expression takes into account the fact that $u_x$ vanishes in the peripheral region, while $\partial_X f_0^p=\partial_X f_0^c=0$, and $\partial_X f_1^c$ vanishes in the central region.  The next-to-leading order components of the effective diffusivity read:
\begin{align}
&D_A= \frac{2}{\langle h^2\rangle}\int_{-1/2}^{1/2}\dd X    \int_0^{a} dy\,y  [u_x^{(0)}f_1^c +u_x^{(1)}f_0^c ], \label{eq:DA3D} \\
&D_B=-\frac{2D}{\langle h^2 \rangle} \int_{-1/2}^{1/2}\dd X     \int_{a}^{h(X)}dy \,y \partial_X f_2^p, \\
&D_C=-\frac{2Da}{\langle h^2\rangle} \int_{-1/2}^{1/2}\dd X       \int_{-\infty}^\infty dY [ \partial_X f^*(X,Y)-\theta(Y) ]. 
 \end{align}
Using \cref{eq:Resultf03D,eq:ValueC03D,eq:Resultf1c3D}, the integrals in the leading order term can be calculated, and we find that
\begin{align}
D_e^{(0)}=\frac{ D a}{a+2\delta}+ \frac{U^2a}{192 D(a+2\delta)^3}    \Bigg\{
 a^4+12a^3\delta\nonumber\\
 +44a^2\delta^2 + 96 \int_a^{h_m}dy\frac{\left(\int_y^{h_m}dy' W(y')y' \right)^2}{W(y)y}  \Bigg\},
\end{align}
where we have defined $\delta$ via the formula
\begin{align}
\langle h^2\rangle=a^2+ 2 a\delta. 
\end{align}

Next, the term $D_A$, defined in \cref{eq:DA3D}, can be calculated by using the previously found expressions for $f_n$, leading to
\begin{align}
D_A= \beta U^2& \frac{a^3 \delta + 8 a^2 \delta^2 + 24 \int_a^{h_m}dy\frac{\left(\int_y^{h_m}dy' W(y')y' \right)^2}{W(y)y}  ) }{12D (a + 2 \delta)^3}\nonumber\\
&-\frac{v_0}{\langle h^2 \rangle}\langle X h^2(X)\rangle. 
\end{align} 
Now, we evaluate the term $D_B$, which reads
\begin{align}
D_B=&-\frac{2D}{\langle h^2 \rangle} \int_{-1/2}^{1/2}\dd X     \int_{a}^{h(X)}dy \, y \partial_X f_2^p \nonumber\\
=&-\frac{2}{\langle h^2 \rangle} \int_{-1/2}^{1/2}dX     \int_{a}^{h(X)}dy \,y \nonumber\\ \Bigg\{&\left[v_0-D\frac{1}{y}\partial_y (y \partial_y f_0^p(y))\right]X 
+\frac{1}{y}\partial_yG_2^p(y)\Bigg\},
\end{align}
where we have used \cref{eq:f2p3D}. Integrating over $y$ and noting that $\int_{-1/2}^{1/2}dX \, X=0$ we obtain
\begin{align}
D_B=&-\frac{2}{\langle h^2 \rangle} \int_{-1/2}^{1/2}dX
\Bigg\{  G_2^p(h(X))-G_2^p(a)+\nonumber\\ + &\left(  \frac{v_0}{2}h^2(X) -D   h(X) (f_0^p)'(h(X))  \right)X \Bigg\}.
\end{align}
We recall \cref{eq:054923D}, 
\begin{align}
& D X h(X)(f_0^p)'(h(X)) =\nonumber\\
&v_0 \int_{-1/2}^X \dd w\, h'(w)h(w)  w + G_2^p(h(X)).  \label{eq:054923Dbis}
\end{align}
Using this equation, we obtain
\begin{align}
D_B=&-\frac{2}{\langle h^2 \rangle} \int_{-1/2}^{1/2}\dd X \Bigg\{      \frac{v_0}{2}  h^2(X)     X \nonumber\\
&-G_2^p(a)-v_0 \int_{-1/2}^X \dd w\,  h'(w)h(w)  w\Bigg\}.
\end{align}
Writing \cref{eq:054923Dbis} for $X=-1/2$ and $X=1/2$, we can show that
\begin{align}
G_2^p(a)= -\frac{v_0}{2} \int_{-1/2}^{1/2} \dd w \, h'(w)h(w)  w. 
\end{align}
Next, using integrations by parts, we have
\begin{align}
 &\int_{-1/2}^{1/2}\dd X \int_{-1/2}^X \dd w\, h'(w)h(w)  w\nonumber\\
 &= -\int_{-1/2}^{1/2}\dd X X^2   h'(X)h(X)  +  \frac{1}{2} \int_{-1/2}^{1/2} \dd w \, h'(w)h(w)  w\nonumber\\
 &=\int_{-1/2}^{1/2}\dd X (2X)   \frac{h^2(X)-a^2}{2}   + \frac{1}{2}  \int_{-1/2}^{1/2} \dd w \, h'(w)h(w)  w. 
 \end{align}
 With these arguments we finally obtain
 \begin{align}
 D_B=&\frac{v_0}{\langle h^2 \rangle}\langle X h^2(X)\rangle. 
 \end{align}
Last, the term $D_C$ is almost the same as in the 2D case, and is given by
\begin{align}
D_C=\frac{Da \,2\ln 2}{\pi \, \langle h^2\rangle }.
\end{align}

We define $\tau$ as the mean escape time out of a protrusion. Using the formulas of Ref.~\cite{szabo1980first}, we identify this time as
\begin{align}
\tau=&k_d^{-1}=\frac{2}{D\langle h^2-a^2\rangle}\int_a^{h_m}dy\frac{\left(\int_y^{h_m}dy' W(y')y' \right)^2}{W(y)y}\nonumber\\
&=\frac{1}{D\delta a}\int_a^{h_m}dy\frac{\left(\int_y^{h_m}dy' W(y')y' \right)^2}{W(y)y}.
\end{align}
Using this value and the obtained expressions for $D_e^{(0)},D_A,D_B,D_C$, we conclude that the expression of the effective diffusivity in a strongly corrugated axisymmetric channel is given by the following formula, valid at next to leading order for small $L$:
\begin{align}
D_e=&\frac{ D a}{a+2\delta}+  \frac{DL 2\ln 2}{\pi \, (a+2\delta) }\nonumber\\
&+\frac{U^2a^2\left(
 a^3+12a^2\delta+44a\delta^2 + 96  D\tau \delta \right)}{192 D\,(a+2\delta)^3}    \nonumber\\
& +\beta L a U^2 \frac{a^2 \delta + 8 a \delta^2 + 24 D\delta\tau  }{12D \,(a + 2 \delta)^3}\label{eq:De3Dn}.
\end{align}
This agrees with the results of Ref.~\cite{levesque2012taylor} when one identifies $\delta=k_a/k_d$ (when one neglects the slip velocity).

\end{document}